\documentclass[reqno,12pt,a4paper]{article}

\usepackage{amsfonts,amssymb,amsthm, 
amsmath,amscd, bbm, bm, mathabx, 
mathrsfs,delarray,subfigure}
\usepackage[dvipsnames]{xcolor}
\usepackage{hyperref,cite}
\usepackage{xypic}
\usepackage{sectsty}

\usepackage{accents}

\sectionfont{\fontsize{12}{15}\selectfont}
\subsectionfont{\fontsize{12}{15}\selectfont}

\usepackage{enumitem}

\DeclareMathOperator{\ad}{ad}

\DeclareMathOperator{\id}{id}

\DeclareMathOperator{\Fun}{Fun}

\DeclareMathOperator{\Vect}{Vect}

\DeclareMathOperator{\End}{End}

\DeclareMathOperator{\grtr}{grtr}

\DeclareMathOperator{\ee}{e}

\DeclareMathOperator{\CE}{CE}

\DeclareMathOperator{\dvg}{div}
\DeclareMathOperator{\crit}{crit}

\numberwithin{equation}{subsection} 
\numberwithin{subsection}{section} 

\newcommand{\ceqref}[1]{{\textcolor{blue}{\eqref{#1}}}}
\newcommand{\cref}[1]{{\textcolor{blue}{\ref{#1}}}}
\newcommand{\ccite}[1]{{\textcolor{blue}{\!\cite{#1}}}}

\newcommand{\ppp}{{\hbox{$\prod$}}}

\newcommand{\ul}[1]{{\underline{#1}}}

\newcommand{\ms}{\scriptscriptstyle}
\newcommand{\bcdot}{{\ms\bullet}}

\newcommand{\mathsans}[1]{{{\sf #1}}}

\font\euler=eusm10 at 12.8 truept
\font\scripteuler=eusm7
\font\scriptscripteuler=eusm5 
\newtheorem{defi}{{\sf Definition}}[section]
\newtheorem{prop}{{\sf Proposition}}[section]

\newtheorem{lemma}{{\sf Lemma}}[section]

\begin{document}

\vskip1.5cm
\begin{large}
{\flushleft\textcolor{blue}{\sffamily\bfseries Exact renormalization group in Batalin--Vilkovisky theory}}  
\end{large}
\vskip1.3cm
\hrule height 1.5pt
\vskip1.3cm
{\flushleft{\sffamily \bfseries Roberto Zucchini}\\
\it Dipartimento di Fisica ed Astronomia,\\
Universit\`a di Bologna,\\
I.N.F.N., sezione di Bologna,\\
viale Berti Pichat, 6/2\\
Bologna, Italy\\
Email: \textcolor{blue}{\tt \href{mailto:roberto.zucchini@unibo.it}{roberto.zucchini@unibo.it}}, 
\textcolor{blue}{\tt \href{mailto:zucchinir@bo.infn.it}{zucchinir@bo.infn.it}}}


\vskip.5cm
\vskip.5cm 
{\flushleft\sc
Abstract:} 
In this paper, inspired by the Costello's seminal work \ccite{Costello:2007ei}, 
we present a general formulation of exact renormalization group (RG) within the Batalin--Vilkovisky 
(BV) quantization scheme. In the spirit of effective field theory, the BV bracket and Laplacian 
structure as well as the BV effective action (EA) depend on an effective energy scale.
The BV EA at a certain scale satisfies the BV quantum master equation at that scale. The RG flow 
of the EA is implemented by BV canonical maps intertwining the BV structures at different scales. 
Infinitesimally, this generates the BV exact renormalization group equation (RGE). We show that BV RG theory 
can be extended by augmenting the scale parameter space $\mathbb{R}$ to its shifted tangent bundle 
$T[1]\mathbb{R}$. The extra odd direction in scale space allows for a BV RG supersymmetry that 
constrains the structure of the BV RGE bringing it to Polchinski's form \ccite{Polchinski:1983gv}.
We investigate the implications of BV RG supersymmetry in perturbation theory. Finally, we illustrate 
our findings by constructing free models of BV RG flow and EA exhibiting RG supersymmetry 
in the degree $-1$ symplectic framework and studying the perturbation theory thereof. 
We find in particular that the odd partner of effective action describes perturbatively 
the deviation of the interacting RG flow from its free counterpart. 
\vskip.2cm
\par\noindent
MSC: 81T13 81T20 81T45  

\vfil\eject

\tableofcontents

\vfil\eject

\vfil\eject

\section{\textcolor{blue}{\sffamily Introduction}}\label{sec:intro}

\vspace{-1.2mm}

The Batalin--Vilkovisky (BV) quantization scheme \ccite{BV1,BV2} is to date the most powerful 
method for quantizing general classical field theories characterized by possibly non involutive and non 
freely acting gauge symmetry. It has found applications in ordinary gauge theory as well as 
supergravity and string theory. Furthermore, since it proceeds in a functional framework, it 
makes the powerful functional integration methods available. See ref. \ccite{Gomis:1994he} for a readable 
introduction to the subject. 

In BV theory, ghost fields are added to the classical fields to make 
Becchi--Rouet--Stora--Tyutin (BRST) symmetry manifest 
from the onset. Each field is then adjoined by an antifield of opposite statistics.
In this way, the total field space is structured as an odd phase space, with fields and 
antifields playing the role of canonically conjugate coordinates and momenta, respectively. 
Hamiltonian notions such as those 
of BV bracket, an odd analog of the Poisson bracket, and canonical map can in this way 
be formulated and used while keeping manifest covariance and BRST invariance. 

The introduction of the antifields doubles the field content. Gauge fixing roughly 
speaking amounts to the restoration of the original field number and is achieved by restricting to a 
suitable Lagrangian submanifold of the total field space. Independence of the quantum theory 
from the choice of the submanifold requires the BV master action (MA) to satisfy the 
BV quantum master equation (ME). This involves an odd second order differential operator called 
BV Laplacian in addition to the BV bracket. Both the MA and the Laplacian 
are affected by UV divergences which need to be suitably regularized and renormalized. 
Violations of the ME result in gauge anomalies.

The BV quantization procedure, when properly implemented, ensure the existence of well-defined 
propagators making the usual perturbative Feynman graph methods usable in a manifestly BRST 
symmetric framework. 

The invariant physical content of a quantum field theory is encoded in the partition function 
with background sources. The renormalization group (RG) \ccite{Kadanoff:1966wm, Wilson:1974mb} 
is a powerful field theoretic construction aimed to probing such content below a certain energy scale, 
though the content itself is insensitive to the presence of the scale, 
which serves as a mere analysis device. There are a priori
many ways of injecting a scale in a field theory and correspondingly 
there are many versions of the RG equations (RGE). In this paper, we shall concentrate on the so called exact 
RG \ccite{Polchinski:1983gv,Wetterich:1992yh} (to be called simply RG in the following). 
See ref. \ccite{Rosten:2010vm} for an updated review. 

The theory of RG flow and effective action (EA) has been studied in a BV framework 
both from a physical \ccite{Igarashi:2001mf,Igarashi:2001ey} and a mathematical
\ccite{Costello:2007ei,Costello:2011ams,Costello:2016faqft,Gwilliam:2016faqft,Mnev:2008sa} point of view
and in a variety of guises. 
Recent mathematical studies close in spirit to our contribution 
are \ccite{Li:2016gcb,Elliott:2017ahb,Li:2017bv}. 

The study of the BV RG is of considerable interest for its applications 
to relevant theoretical problems. We mention here a few. 
In ref. \ccite{Zwiebach:1992ie}, closed string field theory is quantized using the BV 
algorithm and the BV MA is studied in depth. A family of consistent string 
vertexes is obtained parametrized by a stub length working as an ultraviolet cutoff. 
A renormalization group equation for string field theory is yielded in this way.  
In ref. \ccite{Costello:2012cy}, Kodaira--Spencer theory of gravity \ccite{Bershadsky:1993cx} 
is quantized in the BV RG framework of ref. \ccite{Costello:2007ei} and the genus expansion of the B--model 
is related to the perturbative expansion of the resulting quantum field theory.
In ref. \ccite{Gwilliam:2017axm}, using again the BV RG framework  of ref. \ccite{Costello:2007ei}, 
the holomorphic bosonic string is quantized. 

\subsection{\textcolor{blue}{\sffamily Overview of the BV RG framework of this paper}}\label{subsec:doctrine}  

The effectiveness of the BV RGE as a computational scheme and analysis method 
depends on the form of the RG flow, which is a basic input datum of the RGE itself.
However, there is much that can be learned before committing oneself with a specific choice.

In this paper, we formulate a BV theory of the RG flow and RGE from a very general perspective as 
a BV effective field theory (EFT) at a varying energy scale $t$. Our approach is inspired by the 
seminal work by Costello \ccite{Costello:2007ei}, the indebtedness to which we fully acknowledge, 
but it differs from it in key points to be discussed later. 
In an EFT setting, the BV bracket 
$(-,-)_t$ and Laplacian  $\varDelta_t$ depend on $t$. The BV MA $S_t$ also depends on $t$ and satisfies 
the BV quantum ME 
\begin{equation}
\varDelta_tS_t+\frac{1}{2}(S_t,S_t)_t=0. 
\label{ibvfix7}
\end{equation}
$S_t$ can now be viewed as a BV EA. The RG flow of $S_t$ must be such to preserve the ME \ceqref{ibvfix7}.
The essentially only natural way this can come about is through a group $\varphi_{t,s}$ of 
BV canonical maps, whose pull--back action transforms isomorphically the BV bracket and Laplacian at
any scale $s$ into those at another scale $t$. The ME remains satisfied identically in $t$ 
if the EA $S_t$ flows as $S_t=\varphi_{t,s}{}^*S_s+r_{\varphi t,s}$, where 
$r_{\varphi t,s}=\frac{1}{2}\ln J_{\varphi_{t,s}}$ is the logarithmic Jacobian of $\varphi_{t,s}$.
Infinitesimally, then, $S_t$ obeys an RGE of the form  
\begin{equation}
\frac{dS_t}{dt}=\varphi^\bcdot{}_tS_t+r^\bcdot{}_{\varphi t}, 
\label{ibvfix19}
\end{equation}
in which the first and second term of the right hand side can be identified with the familiar 
``classical'' and ``quantum'' contributions to the RG flow. 
We justify this approach to the RG by a biased revisitation of the BV quantization scheme
in sect. \cref{sec:bvfunintren}, where the interplay of the RG flow and gauge fixing is also 
discussed.  

In  sect. \cref{sec:rgsusy}, relying only on the general properties shared by every RG flow, we explore 
the possibility of using as energy scale space more general manifolds than just the real line
$\mathbb{R}$. The fact that BV theory belongs to the realm of graded algebra and geometry
indicates that the scale space may be promoted to a graded manifold. 
We have found particularly useful to employ the shifted tangent space $T[1]\mathbb{R}$ of 
$\mathbb{R}$ rather $\mathbb{R}$ alone for this role. We call this RG set--up ``extended''
to distinguish it from the customary ``basic'' set--up. $T[1]\mathbb{R}$ is coordinatized 
by a degree $0$ parameter $t$, to be identified with the usual RG energy scale, and a further degree $1$ parameter 
$\theta$. In the extend set--up, so, the BV bracket $(-,-)_{t\theta}$ and Laplacian  
$\varDelta_{t\theta}$ depend on both $t$ and $\theta$. The BV MA $S_{t\theta}$ also depends on $t$, $\theta$
and satisfies the extended version of the BV ME, 
\begin{equation}
\varDelta_{t\theta}S_{t\theta}+\frac{1}{2}(S_{t\theta},S_{t\theta})_{t\theta}=0. 
\label{ibveffact25/1}
\end{equation}
$S_{t\theta}$ is the extended BV EA. In analogy to basic case, its RG flow is governed by a group 
$\varphi_{t\theta,s\zeta}$ of canonical maps relating the BV bracket and Laplacian at any extended 
scale $s,\zeta$ to those at another scale $t,\theta$ according to the law 
$S_{t\theta}=\varphi_{t\theta,s\zeta}{}^*S_{s\zeta}+r_{\varphi t\theta,s\zeta}$. At the infinitesimal level, 
this entails a more structured RGE than \ceqref{ibvfix19}. Expanding in powers of the 
odd parameter $\theta$, we have  $(-,-)_{t\theta}=(-,-)_t\pm\theta (-,-)^\star{}_t$
and $\varDelta_{t\theta}=\varDelta_t+\theta \varDelta^\star {}_t$, where $(-,-)^\star{}_t$
is a degree $0$ graded symmetric bracket and $\varDelta^\star {}_t$ is a degree $0$ Laplacian. 
Further, $S_{t\theta}=S_t+\theta S^\star{}_t$, where $S_t$ is to be identified with the usual BV EA and $S^\star{}_t$
is a degree $-1$ partner of it. Then, the extended RGE expresses $S^\star{}_t$ in terms of $S_t$ and 
and yields an RGE for $S_t$ of the form 
\begin{equation}
\frac{dS_t}{dt}=\varDelta^\star{}_tS_t+\frac{1}{2}(S_t,S_t)^\star{}_t
+\bar\varphi^\bcdot{}_tS_t+\bar r^\bcdot{}_{\varphi t},  
\label{ibveffact27}
\end{equation}
where the last two terms in the right hand side are ``seed'' terms. In this way, we obtain a 
version of the BV RGE of the distinctive form of Polchinski's \ccite{Polchinski:1983gv} by purely
algebraic and geometric means.  By adjoining the RG scale with an odd partner, we have 
further revealed the existence of an  intriguing sort of RG supersymmetry with potential
implications for perturbation theory. 

Since presently we have no proof that the basic RG set--up can be embedded in the extended one in 
general, it is important to work out non trivial models in which the extended set--up is implemented. 
In  sect. \cref{sec:models}, working in the $\mathfrak{gl}(1|1)$ degree $-1$ 
symplectic framework originally developed by Costello in ref. \ccite{Costello:2007ei}, 
we illustrate a free model of BV RG flow and EA in the 
extended set--up. We obtain explicit expressions for the free EA $S^0{}_t$ and its odd partner 
$S^{0\star}{}_t$. We then investigate the associated perturbation theory, where the full EA
and partner thereof $S_t$, $S^\star{}_t$ can be expanded as
\begin{equation}
S_t=S^0{}_t+I_t, \qquad S^\star{}_t=S^{0\star}{}_t+I^\star{}_t,
\label{}
\end{equation}
$I_t$, $I^\star{}_t$ being interaction terms expressed as formal power series of $\hbar$. 
We recover the ME and RGE of $I_t$ originally obtained by Costello and obtain a further ME involving 
simultaneously $I_t$, $I^\star{}_t$. We also show that, through this latter, the RGE of $I_t$ 
can be cast in a form analogous to Polchinsi's. Finally, we find that the odd interaction action partner
$I^\star{}_t$ describes perturbatively the deviation of the interacting RG flow from its free counterpart. 

In the final sect. \cref{sec:beyond}, we speculate on possible ramifications and applications 
of our results in mathematical physics, geometry and topology. 

The present paper aims to present the field theoretic foundations of our BV formulation  
of RG theory. In a more mathematical oriented paper \ccite{Zucchini:2017ip}, we shall 
reformulate the results obtained here in the framework of the abstract theory of BV algebras and manifolds. 

\subsection{\textcolor{blue}{\sffamily Relation to Costello's BV RG theory}}\label{subsec:relation} 

In spite of evident formal similarities, there is a basic difference between the RG 
framework  propounded in the present paper and Costello's \ccite{Costello:2007ei}.
Costello's approach is Wilsonian in nature \ccite{Wilson:1974mb}. In the 
familiar formulation of Wilson's RG, the EFT at the energy scale $t$
is described by an EA $S_t$ and functional integration is restricted to 
the finite dimensional subspace $\mathcal{F}_t$ of field space $\mathcal{F}$ 
spanned by the field modes of energy up to $t$. The RG flow is determined by 
the requirement that the partition function is independent from $t$. Since 
 flow to a lower energy scale 
involves a truncation of the functional integration domain $\mathcal{F}_t$ to
a proper subspace, the flow cannot be implemented by invertible field maps.
The approach set forth by us is instead Polchinskian \ccite{Polchinski:1983gv}. 
In Polchinski's RG framework, the EFT at the energy scale $t$ is described by an EA $S_t$ as in Wilson's case 
but functional integration is extended to the full field space $\mathcal{F}$. 
The restriction of integration to the field modes up to the scale $t$ 
is due to the special dependence of $S_t$ on the energy 
scale $t$ which through the Boltzmann weight $\ee^{S_t}$ causes an
exponential damping of the contribution of the field modes of energy 
above $t$ in the functional integral. The RG flow is determined again 
by the requirement that the partition function is independent from $t$,
but proceeds through invertible field maps \ccite{Latorre:2000qc,Latorre:2000jp,Sumi:2000xp}. 
(Above, we have ignored the complications related to gauge symmetries and gauge fixing 
for the sake of the argument.) 

Wilson and Polchinski RG are presumably related but it is difficult formulate 
this relationship in precise terms. In a Polchinskian perturbative RG framework, 
the restriction to energy modes below the scale $t$ is due as a rule 
to a modification of the standard kinetic operator $H$ of the free action $S^0{}_t$ into $H/k(H/t)$,
where $k(x)$ is some rapidly decreasing cut--off function, e. g. $\ee^{-H/t}$. This 
suppresses the field modes of energy greater than $t$ through the Boltzmann weight 
$\ee^{S^0{}_t}$. The Wilsonian perturbative RG framework would roughly correspond to replace $k(x)$ with 
a function taking the value $1$ for $x<1$ and $0$ for $x>1$. 
Wilson and Polchinski RG are therefore akin but distinct. So, are Costello's and this paper's 
BV RG formulations. This also impinges on the way beta functions are
defined in the two approaches, though this matter is not dealt with in this paper. 

\subsection{\textcolor{blue}{\sffamily Strengths and weaknesses of the BV RG framework presented}}\label{subsec:limitation} 

The BV RG formulation proposed in the present study, relying heavily on functional integral 
manipulations, is expectedly difficult to put on a firm rigorous ground. E. g., the quantum 
part of the RG equation is related to the Jacobian of an invertible field map which as a rule requires 
regularization. Yet, in quantum field theory it is normally possible to cope with this kind of problems 
relying also on the indications provided by finite dimensional analog models. 

Our approach to the BV RG is based on BV geometry. Most studies on this subject available in the 
literature, including  A. Schwarz's foundational work \ccite{Schwarz:1992nx}, assume a finite 
dimensional context. Even in this simplified form they retain however
a considerable interest for the clues and intuition they provide. This is the point of view
taken in this article.

Most of what is stated in the following is therefore strictly true in finite dimensions.
A part of analysis is admittedly conjectural, although we feel that the assumptions
we made along the way are natural and neatly fit into the whole picture. The main challenge lying 
ahead is testing the theory in non trivial field theoretic models dealing with 
the difficult technical problems arising in infinite dimensions. We leave this work for the future. 

Costello's BV RG theory \ccite{Costello:2007ei} is presently on a firm mathematical ground.
With this study, we certainly do not claim having built an alternative BV RG theory of comparable 
strength and soundness, but only explore an alternative. The exact RG philosophy we espouse
is in fact widely used by quantum field theorists interested in concrete physical applications.

\vfil\eject

\section{\textcolor{blue}{\sffamily Renormalization group in Batalin--Vilkovisky theory}}\label{sec:bvfunintren}

In this section, we shall first review briefly the foundations of BV theory from a geometric perspective.  
This will also serve to set our terminology and notation. 
We shall then expound our BV formulation of RG theory, its interplay with gauge fixing and 
its implications for perturbation theory. What is stated below is strictly true 
only in a finite dimensional setting. In the infinite dimensional context of quantum field theory that 
interest us, our treatment is merely formal. Most of the geometrical theory was developed in Schwarz's seminal work 
\ccite{Schwarz:1992nx}. See refs. \ccite{Cattaneo:2010re,Mnev:2017oko} for a recent up to date review.


\subsection{\textcolor{blue}{\sffamily Elements of  BV  geometry}}\label{subsec:bvint}

In this subsection, we review briefly the basic notions of BV geometry highlighting those 
points which will be relevant in the subsequent applications.  

BV geometry is the theory of BV manifolds. A BV manifold $\mathcal{M}$ 
is a $\mathbb{Z}$--grad\-ed manifold endowed with  a degree $-1$ symplectic form 
$\omega_{\mathcal{M}}$ and a Berezinian $\mu_{\mathcal{M}}$, called BV form and 
 measure respectively, obeying a compatibility condition. 

Canonically associated with the BV form $\omega_{\mathcal{M}}$ is a bilinear bracket 
$(-,-)_{\mathcal{M}}:\Fun(\mathcal{M})\times\Fun(\mathcal{M})\rightarrow\Fun(\mathcal{M})$, called BV bracket,
as follows. For every function $f\in\Fun(\mathcal{M})$, there is a unique vector field
$\ad_{\mathcal{M}} \hspace{-1pt}f\in\Vect(\mathcal{M})$ such that 
\begin{equation}
i_{\ad_{\mathcal{M}} \hspace{-1pt}f}\omega_{\mathcal{M}}=df.
\label{bvint1}
\end{equation}
Then, for any function pair $f,g\in\Fun(\mathcal{M})$, one has 
\begin{equation}
(f,g)_{\mathcal{M}}=\ad_{\mathcal{M}}\hspace{-1pt}fg.
\label{bvint2}
\end{equation}
$(-,-)_{\mathcal{M}}$ is a Gerstenhaber bracket, that is it has degree $1$, is shifted graded antisymmetric, 
obeys the shifted graded Jacobi identity and acts as a shifted graded derivation 
on both its arguments. 

The BV measure $\mu_{\mathcal{M}}$ induces a linear integration map
$\int_{\mathcal{M}}\mu_{\mathcal{M}}\hspace{.5pt}-:\Fun(\mathcal{M})$ $\rightarrow\mathbb{C}$, called BV integral.
$\int_{\mathcal{M}}\mu_{\mathcal{M}}\hspace{.5pt}-$ has degree $\dim_{\mathrm{virt}}\mathcal{M}$,  
the sum of the degrees of the even minus the sum of the degrees of the odd coordinates 
of ${\mathcal{M}}$. There is a linear operator $\dvg_{\mathcal{M}}:\Vect(\mathcal{M})\rightarrow \Fun(\mathcal{M})$ 
associated with $\mu_{\mathcal{M}}$, called BV di\-vergence. 
$\dvg_{\mathcal{M}}$ is the degree $0$ first order differential operator 
uniquely defined by the property that \hphantom{xxxxxxxxxxxxxxx}
\begin{equation}
\int_{\mathcal{M}}\mu_{\mathcal{M}} \hspace{.5pt}Xf=-\int_{\mathcal{M}}\mu_{\mathcal{M}}\dvg_{\mathcal{M}} Xf.
\label{bvint3}
\end{equation}
for any vector field $X\in\Vect(\mathcal{M})$ and function $f\in\Fun(\mathcal{M})$. 

Combining the BV form and measure, one can construct the BV Laplacian, the linear 
operator $\varDelta_{\mathcal{M}}:\Fun(\mathcal{M})\rightarrow\Fun(\mathcal{M})$ given by 
\begin{equation}
\varDelta_{\mathcal{M}}f=\frac{(-1)^{|f|}}{2}\dvg_{\mathcal{M}}\ad_{\mathcal{M}} f
\label{bvint4}
\end{equation}
for $f\in\Fun(\mathcal{M})$. $\varDelta_{\mathcal{M}}$ is a degree $1$ second order differential operator. 
The com\-patibility of the BV form $\omega_{\mathcal{M}}$ and measure $\mu_{\mathcal{M}}$ consists in 
$\varDelta_{\mathcal{M}}$ being nilpotent, 
\begin{equation}
\varDelta_{\mathcal{M}}{}^2=0.
\label{bvint5}
\end{equation}
This is a non trivial constraint relating $\omega_{\mathcal{M}}$ and $\mu_{\mathcal{M}}$. 

The BV Laplacian $\varDelta_{\mathcal{M}}$  enjoys a number of relevant properties. 
$\varDelta_{\mathcal{M}}$ encodes the BV bracket $(-,-)_{\mathcal{M}}$ as 
\begin{equation}
(f,g)_{\mathcal{M}}=(-1)^{|f|}(\varDelta_{\mathcal{M}}(fg)-\varDelta_{\mathcal{M}}fg
-(-1)^{|f|}f\varDelta_{\mathcal{M}}g), \vphantom{\bigg]}
\label{bvint6}
\end{equation}
for $f,g\in\Fun(\mathcal{M})$. 
Furthermore, $\varDelta_{\mathcal{M}}$ obeys standard integral identities such as the BV Green identity  
\hphantom{xxxxxxxxxx}
\begin{equation}
\int_{\mathcal{M}}\mu_{\mathcal{M}} (\varDelta_{\mathcal{M}} fg-(-1)^{|f|}f\varDelta_{\mathcal{M}} g)=0
\label{bvint7}
\end{equation}
and the BV Laplace identity \hphantom{xxxxxxxx}
\begin{equation}
\int_{\mathcal{M}}\mu_{\mathcal{M}}\hspace{.5pt}\varDelta_{\mathcal{M}} f=0,
\label{bvint8}
\end{equation}
where $f,g\in\Fun(\mathcal{M})$ are any functions. 

A Lagrangian submanifold $\mathcal{L}$ of $\mathcal{M}$, is a maximal submanifold of $\mathcal{M}$ 
such that $i_{\mathcal{L}}{}^*\omega_{\mathcal{M}}=0$, where $i_{\mathcal{L}}$ is the natural injection. 
There exists a Berezinian $\mu_{\mathcal{M}}|_{\mathcal{L}}{}^{1/2}$ on $\mathcal{L}$ 
which is a tensor square root of the restriction of $\mu_{\mathcal{M}}$ to $\mathcal{L}$, \pagebreak 
\begin{equation}
i_{\mathcal{L}}{}^*\mu_{\mathcal{M}}=(\mu_{\mathcal{M}}|_{\mathcal{L}}{}^{1/2})^{\otimes 2}.
\label{bvint14}
\end{equation}
Using $\mu_{\mathcal{M}}|_{\mathcal{L}}{}^{1/2}$, one builds a linear $\mathcal{L}$ --integration  map
$\int_{\mathcal{L}}\mu_{\mathcal{M}}|_{\mathcal{L}}{}^{1/2}\hspace{.5pt}-:\Fun(\mathcal{M})$ $\rightarrow \mathbb{C}$.
A BV version of Stokes' theorem holds. For any function $f\in\Fun(\mathcal{M})$, 
\begin{equation}
\int_{\mathcal{L}}\mu_{\mathcal{M}}|_{\mathcal{L}}{}^{1/2}\varDelta_{\mathcal{M}}f=0.
\label{bvint15}
\end{equation}
Further, if $\mathcal{L}$ and $\mathcal{L}'$ are two Lagrangian submanifolds of $\mathcal{M}$
whose bodies are homologous as cycles of the body of $\mathcal{M}$, then 
\begin{equation}
\int_{\mathcal{L}'}\mu_{\mathcal{M}}|_{\mathcal{L}}{}^{1/2}f=\int_{\mathcal{L}}\mu_{\mathcal{M}}|_{\mathcal{L}}{}^{1/2}f
\label{bvint16}
\end{equation}
for any function $f\in\Fun(\mathcal{M})$ satisfying 
\begin{equation}
\varDelta_{\mathcal{M}}f=0. 
\label{bvint17}
\end{equation}
As the BV Laplacian $\varDelta_{\mathcal{M}}$ is nilpotent by \ceqref{bvint5}, the BV cohomology
$H_{\varDelta_{\mathcal{M}}}{}^*(\mathcal{M})$ is defined. By the BV Stokes' theorem, the $\mathcal{L}$ integration map
factors through 
a map $\int_{\mathcal{L}}\mu_{\mathcal{M}}|_{\mathcal{L}}{}^{1/2}\hspace{.5pt}-:H_{\varDelta_{\mathcal{M}}}{}^*(\mathcal{M})
\rightarrow \mathbb{C}$.

Let $\mathsans{B}$ be a fixed collection of BV manifolds which are diffeomorphic as $\mathbb{Z}$--graded manifolds.
For instance, $\mathsans{B}$ may consist of BV manifolds which are equal as $\mathbb{Z}$--graded manifolds
but have different BV forms and measures, the situation mostly considered in this paper. 

The natural invertible maps $\phi:\mathcal{M}'\rightarrow\mathcal{M}$ of two BV manifolds 
$\mathcal{M}, \mathcal{M}'\in\mathsans{B}$ are the canonical ones, that is those 
with the property that $\phi^*\omega_{\mathcal{M}}=\omega_{\mathcal{M}'}$. 
They preserve the BV bracket, so that  \hphantom{xxxxxxxx}
\begin{equation}
\phi^*(f,g)_{\mathcal{M}}=(\phi^* f,\phi^* g)_{\mathcal{M}'} 
\label{bvint9}
\end{equation}
for all functions $f,g\in\Fun(\mathcal{M})$. 

The BV integral has the usual covariance properties. If $\phi:\mathcal{M}'\rightarrow \mathcal{M}$ 
is a canonical map of two BV manifolds $\mathcal{M}, \mathcal{M}'\in\mathsans{B}$, then there is a 
nowhere vanishing degree $0$ function $J_\varphi\in\Fun(\mathcal{M}')$, called the Jacobian of $\phi$, such that 
\begin{equation}
\int_{\mathcal{M}}\mu_{\mathcal{M}} \hspace{.5pt} f=\int_{\mathcal{M}'}\mu_{\mathcal{M}'} \hspace{.5pt} J_\phi 
\hspace{.5pt} \phi^*  f
\label{bvint10}
\end{equation}
for any $f\in\Fun(\mathcal{M})$. The logarithmic Jacobian 
\begin{equation}
r_\phi=\ln J_\phi{}^{1/2} 
\label{bvint11}
\end{equation} 
is a natural, equivalent substitute for $J_\phi$. 
It obeys the equation 
\begin{equation}
\varDelta_{\mathcal{M}'}r_\phi+\frac{1}{2}(r_\phi,r_\phi)_{\mathcal{M}'}=0.
\label{bvint12}
\end{equation}
A canonical map $\phi$ is said special if $r_\phi=0$. 

The pull--back action of a canonical map $\phi:\mathcal{M}'\rightarrow \mathcal{M}$ of BV manifolds
$\mathcal{M}, \mathcal{M}'\in\mathsans{B}$ does not intertwine between the BV Laplacians $\varDelta_{\mathcal{M}}$, 
$\varDelta_{\mathcal{M}'}$. In fact, 
\begin{equation}
\varDelta_{\mathcal{M}'}\phi^* f-\phi^* \varDelta_{\mathcal{M}}f+(r_\phi, \phi^* f)_{\mathcal{M}'}=0
\label{bvint13}
\end{equation}
for $f\in\Fun(\mathcal{M})$. It does however if $\phi$ is special.
Relation \ceqref{bvint9} expresses the non covariance of $\varDelta_{\mathcal{M}}$ 
for a generic $\phi$. 

The BV manifolds of the collection $\mathsans{B}$ and their canonical maps
constitute a groupoid under composition and inversion, the canonical groupoid of $\mathsans{B}$. 
The BV manifolds of $\mathsans{B}$ together with the special canonical maps
constitute a subgroupoid of the canonical groupoid of $\mathsans{B}$, the special 
canonical groupoid of $\mathsans{B}$. However, we shall not rely on this categorical 
interpretation in what follows. 

If $\phi:\mathcal{M}'\rightarrow \mathcal{M}$ is a canonical map of BV manifolds $\mathcal{M},\mathcal{M}'\in\mathsans{B}$
and $\mathcal{L}'$ is a Lagrangian submanifold of 
$\mathcal{M}'$, then $\phi_*\mathcal{L}'$ is a Lagrangian submanifold of $\mathcal{M}$ as well. In this case, 
\ceqref{bvint14} and the identity $i_{\phi_*\mathcal{L}'}\circ \phi|_{\mathcal{L}'}=\phi\circ i_{\mathcal{L}'}$ imply that 
\begin{equation}
\int_{\phi_*\mathcal{L}'}\mu_{\mathcal{M}}|_{\phi_*\mathcal{L}'}{}^{1/2}\hspace{.5pt}f
=\int_{\mathcal{L}'}\mu_{\mathcal{M}'}|_{\mathcal{L}'}{}^{1/2}\hspace{.5pt}J_\phi{}^{1/2}\hspace{.5pt}\phi^* f
\label{bvint18}
\end{equation}
for $f\in\Fun(\mathcal{M})$. 

A prototypical BV manifold $\mathcal{M}$ is the $-1$ shifted cotangent bundle
$T^*[-1]\mathcal{F}$ \linebreak of a $\mathbb{Z}$--graded manifold $\mathcal{F}$ with canonical BV form
$\omega_{\mathrm{can}}$ and measure $\mu_{\mathrm{can}}$.
$\omega_{\mathrm{can}}$ and $\mu_{\mathrm{can}}$ are given by the familiar relations in terms of the 
base and fiber coordinates $x^a$ and $x^*{}_a$, $\omega_{\mathrm{can}}=dx^*{}_adx^a$ and $\mu_{\mathrm{can}}=
d^{\dim\mathcal{F}}x\hspace{.5pt}d^{\dim\mathcal{F}}x^* $. The BV bracket and Laplacian are in this case
\begin{equation}
(f,g)_{\mathrm{can}}=(-1)^{(|f|+1)\epsilon^a}\partial_af\partial^{* a}g-(-1)^{(|f|+1)\epsilon^*{}_a}\partial^{* a}f\partial_ag
\label{bvint19}
\end{equation}
and \hphantom{xxxxxxxxxxxxxxxxxxxxxxxxxxx} 
\begin{equation}
\varDelta_{\mathrm{can}}f=(-1)^{\epsilon^a}\partial_a\partial^{* a}f, 
\label{bvint20}
\end{equation}
respectively, where $\epsilon^a=|x^a|$, $\epsilon^*{}_a=|x^*{}_a|$. 

Every BV manifold $\mathcal{M}$ with BV form 
$\omega$ and measure $\mu$ admits locally special Darboux coordinates $x^a$ and $x^*{}_a$
such that $\omega$ and $\mu$ are expressed as $\omega_{\mathrm{can}}$ and $\mu_{\mathrm{can}}$ 
in terms $x^a$ and $x^*{}_a$. Thus, the local structure of BV manifold $\mathcal{M}$ 
is  that of a prototypical BV manifold $T^*[-1]\mathcal{F}$ for some $\mathcal{F}$.


\subsection{\textcolor{blue}{\sffamily BV quantization scheme}}\label{subsec:bvfix}

In this subsection, we review the BV quantization algorithm or BV theory for short. 
BV geometry, as formulated in subsect. \cref{subsec:bvint}, 
informs BV theory and makes its structure intuitive. 

In field theory, one is initially given with a space $\mathcal{F}_{\mathrm{cl}}$ of degree 
$0$ classical fields $\psi_{\mathrm{cl}}{}^i$ and a degree $0$ action functional
$S_{\mathrm{cl}}$ 
invariant under the action of a distribution $\mathcal{G}$ 
of the tangent space $T\mathcal{F}_{\mathrm{cl}}$ of $\mathcal{F}_{\mathrm{cl}}$, 
representing gauge symmetries in a broad sense. 
One is further interested in the computation of quantum correlators of field functionals
$f$ of $\mathcal{F}_{\mathrm{cl}}$ which are invariant under the action of $\mathcal{G}$.
In general, $\mathcal{G}$ is not 
freely acting nor it is involutive, though as a rule it is when restricted to the submanifold of 
$\mathcal{F}_{\mathrm{cl}}$ of fields obeying the Euler--Lagrange equations, that is 
the critical locus $\crit(S_{\mathrm{cl}})$ of $S_{\mathrm{cl}}$ in $\mathcal{F}_{\mathrm{cl}}$. 
As well known, a functional integral approach to quantization in a situation of this type
is problematic. BV theory offers an elegant solution for this difficulty. 

The implementation of the BV quantization algorithm involves the following four steps.
First, one enlarges the field space $\mathcal{F}_{\mathrm{cl}}$ by adding to the classical 
fields a ghost field of suitable positive degree for each of the independent generators 
of the distribution $\mathcal{G}$. One obtains in this way a space $\mathcal{F}$ of fields 
$\psi^a$ of non negative degree. Second, 
one adjoins a negatively graded antifield $\psi^*{}_a$ to each field 
$\psi^a$ such that the degrees of $\psi^*{}_a$ and  $\psi^a$  add up to $-1$. 
The resulting total field space is structured as the $-1$ shifted cotangent bundle
$T^*[-1]\mathcal{F}$ of $\mathcal{F}$ with fields and antifields 
parametrizing respectively its base and fibers. $T^*[-1]\mathcal{F}$ is then endowed 
with the canonical BV form and measure $\omega_{\mathrm{can}}$ and $\mu_{\mathrm{can}}$, 
rendering it a BV manifold. 
Third, the action $S_{\mathrm{cl}}$ is extended to an action $S$, called BV MA, 
defined on the whole
total field space $T^*[-1]\mathcal{F}$ and such that $S$ reduces to $S_{\mathrm{cl}}$ when 
all the fields and antifields are set to zero except for the original classical fields 
$\psi_{\mathrm{cl}}{}^i$. Fourth, quantization is implemented by
restricting the functional integration to a suitable 
Lagrangian submanifold $\mathcal{L}$ of $T^*[-1]\mathcal{F}$ with the Boltzmann weight $\ee^S$ inserted.
The unnormalized correlator of a functional $f$ on $T^*[-1]\mathcal{F}$ is therefore given by an integral 
of the form \hphantom{xxxxxxxxxxxxx} 
\begin{equation}
Z_S(f)=\int_{\mathcal{L}}\mu_{\mathrm{can}}|_{\mathcal{L}}{}^{1/2} \ee^S f.
\label{bvfix1}
\end{equation}   

$\mathcal{L}$ must be carefully chosen in order to avoid the usual diseases associated with the 
gauge symmetry of $S_{\mathrm{cl}}$. Except for the simplest cases, taking $\mathcal{L}$ 
to be the zero section of $T^*[-1]\mathcal{F}$, corresponding to setting all the antifields 
to zero, will not work. In general, there are infinitely many choices of $\mathcal{L}$. 
Consistency requires that the value of quantum correlators is independent from the choice made.
This entails a restriction on the form of the BV MA $S$: the BV quantum ME 
\begin{equation}
\varDelta_{\mathrm{can}}S+\frac{1}{2}(S,S)_{\mathrm{can}}=0
\label{bvfix2}
\end{equation}
must be obeyed by $S$. 
It also entails a restriction on the allowed functionals $f$ which can be inserted in correlators; 
they must satisfy the equation 
\begin{equation}
\varDelta_{\mathrm{can}S}f=0, 
\label{bvfix3}
\end{equation}
where $\varDelta_{\mathrm{can}S}$ is the degree $1$ second order linear differential operator
defined by 
\begin{equation}
\varDelta_{\mathrm{can}S}f=\varDelta_{\mathrm{can}}f+(S,f)_{\mathrm{can}}. 
\label{bvfix4}
\end{equation}
$\varDelta_{\mathrm{can}S}$ is called covariant BV Laplacian. 
By virtue of \ceqref{bvfix2}, $\varDelta_{\mathrm{can}S}$ is nilpotent 
\begin{equation}
\varDelta_{\mathrm{can}S}{}^2=0 
\label{bvfix5}
\end{equation}
and is therefore characterized by its cohomology.  

Inserting the Boltzmann weight $\ee^S$ in the functional integral \ceqref{bvfix1}  
amounts to redefine the BV measure from $\mu_{\mathrm{can}}$ to 
\pagebreak 
\begin{equation}
\mu_{\mathrm{can}S}=\mu_{\mathrm{can}} \cdot\ee^{2S}. 
\label{bvfix6}
\end{equation}
The BV ME \ceqref{bvfix2} ensures that the measure $\mu_{\mathrm{can}S}$
is compatible with $\omega_{\mathrm{can}}$ as $\mu_{\mathrm{can}}$. 
$\varDelta_{\mathrm{can}S}$ is just the nilpotent BV Laplacian associated 
with $\omega_{\mathrm{can}}$ and $\mu_{\mathrm{can}S}$ according to \ceqref{bvint4}. 


\subsection{\textcolor{blue}{\sffamily BV RG theory}}\label{subsec:bvrenorm}

In a concrete setting in which the fields in $T^*[-1]\mathcal{F}$ are sections of bundles on a 
space--time manifold, the above functional set--up is plagued by ultraviolet (UV) divergences. 
In fact, the canonical BV form and measure $\omega_{\mathrm{can}}$ and $\mu_{\mathrm{can}}$ 
yield a BV Laplacian $\varDelta_{\mathrm{can}}$ 
containing pairs of functional derivatives acting at the same point of space--time 
and thus ill-defined. In such a situation, the whole theoretical scheme is 
purely formal and so virtually unusable. 
To cure this disease, one has to introduce a very high energy scale 
$t_0$ to regularize the UV divergences and use the methods of EFT 
to describe the quantum field theory at a relevant lower energy scale $t$. 

It is natural to expect that, in an EFT description, the formal 
framework of BV quantization should remain essentially unchanged. 
BV theory and BV geometry that underlies it constitute a delicate structure unlike 
to preserve its efficacy and selfconsistency under severe modifications. 
We assume thus that quantization proceeds much as in the unregularized case 
but with the unregularized quantities replaced by effective ones at the given scale $t$. 
The unregularized BV form and measure $\omega_{\mathrm{can}}$ and $\mu_{\mathrm{can}}$ get in this way 
replaced by an effective BV form $\omega_t$ and measure $\mu_t$ and the unregularized BV bracket 
$(-,-)_{\mathrm{can}}$ and Laplacian $\varDelta_{\mathrm{can}}$ by the associated effective BV 
bracket $(-,-)_t$ and Laplacian $\varDelta_t$. There is a price attached to this:
$\omega_t$, $\mu_t$, $(-,-)_t$ and $\varDelta_t$ are mathematically far more complicated than 
$\omega_{\mathrm{can}}$, $\mu_{\mathrm{can}}$, $(-,-)_{\mathrm{can}}$ and $\varDelta_{\mathrm{can}}$ are. 
Further, since the scale $t$ cannot be assigned any a priori special value, 
we are considering a whole one parameter family of BV manifolds 
$T^*[-1]\mathcal{F}_t$ having $T^*[-1]\mathcal{F}$ as underlying graded manifold
and $\omega_t$ and $\mu_t$ as BV form and measure. 
However, the formal properties of the effective structure are 
exactly the same as those of the unregularized one, 
allowing for an in depth theoretical analysis. 

From subsect. \cref{subsec:bvfix}, quantization of the EFT
involves restriction of integration to a Lagrangian 
submanifold $\mathcal{L}_t$ of $T^*[-1]\mathcal{F}_t$ with insertion of the Boltzmann weight 
$\ee^{S_t}$ of a BV EA $S_t$. Unnormalized correlators 
thus read as 
\begin{equation}
Z_S(f)=\int_{\mathcal{L}_t}\mu_{t}|_{\mathcal{L}_t}{}^{1/2} \ee^{S_t} f
\vphantom{\ul{\ul{\ul{\ul{g}}}}}
\label{bvfix1/r}
\end{equation}   
for any functional $f$ on $T^*[-1]\mathcal{F}$. 
The EA $S_t$ must again satisfy the  BV ME 
\begin{equation}
\varDelta_tS_t+\frac{1}{2}(S_t,S_t)_t=0. 
\label{bvfix7}
\end{equation}
Further, the functionals $f$ inserted in correlators must obey 
\begin{equation}
\varDelta_{tS}f=0, 
\label{bvfix8}
\end{equation}
where $\varDelta_{tS}$ is the effective covariant Laplacian defined by 
\begin{equation}
\varDelta_{tS}f=\varDelta_tf+(S_t,f)_t. 
\label{bvfix9}
\end{equation}
Again, by virtue of \ceqref{bvfix7}, $\varDelta_{tS}$ is nilpotent 
\begin{equation}
\varDelta_{tS}{}^2=0
\label{bvfix10}
\end{equation}
and thus characterized by its cohomology.

The variation of the Lagrangian submanifold $\mathcal{L}_t$ and the BV EA $S_t$ under 
continuous shifts of the energy scale $t$, the so called RG flow, 
must be such to preserve the whole BV structure. 
In our EFT formulation, it is natural to assume that the flow
is implemented by scale dependent field redefinitions, 
that is invertible maps of the total field space $T^*[-1]\mathcal{F}$. 

The largest most natural family of maps preserving the BV structure
is the set of morphisms of the canonical groupoid of 
the BV manifold family $T^*[-1]\mathcal{F}_t$ which we introduced above. 
In the present modelization, so, the RG flow is given by a two parameter family
of canonical maps $\varphi_{t,s}:T^*[-1]\mathcal{F}_t\rightarrow T^*[-1]\mathcal{F}_s$
satisfying the relations 
\begin{align}
&\varphi_{t,s}\varphi_{u,t}=\varphi_{u,s},
\vphantom{\Big]}
\label{bvfix11}
\\
&\varphi_{s,t}=\varphi_{t,s}{}^{-1},
\vphantom{\Big]}
\label{bvfix12}
\end{align}
\vspace{-1.05cm}\pagebreak
\begin{align}
&\varphi_{s,s}=\id_{T^*[-1]\mathcal{F}}
\vphantom{\Big]}
\label{bvfix13}
\end{align}
for $s,t\in \mathbb{R}$. The basic requirement that the RG flow must obey is that 
the partition function $Z_S(1)$ remains constant along it, so that for any two value $s,t$ of the 
energy scale \hphantom{xxxxxxxxxxxx}
\begin{equation}
Z_S(1)=\int_{\mathcal{L}_t}\mu_t|_{\mathcal{L}_t}{}^{1/2}\ee^{S_t}
=\int_{\mathcal{L}_s}\mu_s|_{\mathcal{L}_s}{}^{1/2}\ee^{S_s}.
\label{bvfix14}
\end{equation}
By the general result \ceqref{bvint18}, one has 
\begin{equation}
\int_{\varphi_{t,s*}\mathcal{L}_t}\mu_s|_{\varphi_{t,s*}\mathcal{L}_t}{}^{1/2}\ee^{S_s}
=\int_{\mathcal{L}_t}\mu_t|_{\mathcal{L}_t}{}^{1/2}\ee^{\varphi_{t,s}{}^*S_s+\ln J_{\varphi_{t,s}}{}^{1/2}}. 
\label{bvfix15}
\end{equation}
A sufficient condition for the identity \ceqref{bvfix14} to hold is that the Lagrangian
submanifolds $\mathcal{L}_s$, $\mathcal{L}_t$ are related by \hphantom{xxxxxxxxx}
\begin{equation}
\varphi_{t,s*}\mathcal{L}_t=\mathcal{L}_s, \vphantom{\bigg]}
\label{bvfix16}
\end{equation}
at least up to body homology, and that the RG flow of the BV EA $S_t$ is driven 
by the flow map $\varphi_{t,s}$, \hphantom{xxxxxxxxxxxx}
\begin{equation}
S_t=\varphi_{t,s}{}^*S_s+r_{\varphi t,s}, \vphantom{\bigg]}
\label{bvfix18}
\end{equation}
where the logarithmic Jacobian $r_{\varphi t,s}=r_{\varphi_{t,s}}$ is defined in \ceqref{bvint11}. 

Relation \ceqref{bvfix18} implies that the RG flow of the BV EA
is governed by the first order differential equation, 
\begin{equation}
\frac{dS_t}{dt}=\varphi^\bcdot{}_tS_t+r^\bcdot{}_{\varphi t}, 
\label{bvfix19}
\end{equation}
where $\varphi^\bcdot{}_t$ and $r^\bcdot{}_{\varphi t}$ are respectively given by 
\begin{align}
&\varphi^\bcdot{}_t =\frac{\partial \varphi_{t,s}{}^*}{\partial t}\Big|_{s=t},
\vphantom{\Big]}
\label{bvfix20}
\\
&r^\bcdot{}_{\varphi t} =\frac{\partial r_{\varphi t,s}}{\partial t}\Big|_{s=t}.
\vphantom{\Big]}
\label{bvfix21}
\end{align}
\ceqref{bvfix19} is the BV RGE in its more general form. It has the standard form
of an RGE, the first and second term in the right hand side corresponding 
to the so--called classical and quantum term, respectively. 

The concrete \pagebreak embodiment of the BV RGE \ceqref{bvfix19} obeyed by the BV 
EA $S_t$ depends on the form of the RG flow map $\varphi_{t,s}$. At first glance, 
\ceqref{bvfix19} appears to be a non homogeneous linear differential equation in $S_t$,
but this is a deception of notation. 
Relation \ceqref{bvfix18} on which \ceqref{bvfix19} 
rests is in fact a condition simultaneously constraining and relating $\varphi_{t,s}$ and $S_t$. 
The flow map $\varphi_{t,s}$ therefore is not a datum independent from the action $S_t$ 
but in general contains built in information on the latter. The realizations of $\varphi_{t,s}$ 
encountered in concrete applications do indeed depend explicitly on $S_t$. The RGE 
\ceqref{bvfix19}, so, normally is nonlinear in $S_t$.  

The BV RG framework we have just outlined fits with the Polchinski's RG theory \ccite{Polchinski:1983gv} reviewed in subsect. 
\cref{subsec:doctrine} and is also in line with the RG formulations of refs. 
\ccite{Latorre:2000qc,Latorre:2000jp,Sumi:2000xp}, although at this point
we have not yet committed ourselves with any specific assumption about the form of the EA $S_t$.


\subsection{\textcolor{blue}{\sffamily Derived BV EAs \sffamily }}\label{subsec:centext}

In the BV theory of the RG, it is often useful to consider parametrized families of BV EA $S_t$
whose RG flow is governed by a fixed set of flow maps $\varphi_{t,s}$. This is most naturally 
done using the methods of derived geometry along the lines of ref. \ccite{Costello:2016faqft}. 
Here, we shall present a simplified formulation tailored for a functional integral formulation
of BV theory.

In the derived approach, a family of functionals on field space $T^*[-1]\mathcal{F}$ 
is modelled as a functional on $T^*[-1]\mathcal{F}$ valued in a differential graded commutative 
algebra $\mathcal{A}$, which must be thought of as the algebra of functions on a fiducial 
parameter space. The family might depend on the RG scale $t$, while $\mathcal{A}$ itself does not. 
The BV Laplacians $\varDelta_t$, brackets $(-,-)_t$ and RG flow maps $\varphi_{t,s}$ are assumed to be given. 
The $\varDelta_t$, $(-,-)_t$ and the pull--backs $\varphi_{t,s}{}^*$ of the $\varphi_{t,s}$ 
act on functional families through their $\mathcal{A}$--linear extensions. 
A derived BV EA is then a $t$ dependent degree $0$ $\mathcal{A}$--valued 
functional $S_{\mathcal{A}t}$ on $T^*[-1]\mathcal{F}$ defined modulo $\mathcal{A}$--valued constants 
and obeying the derived BV quantum ME 
\begin{equation}
d_{\mathcal{A}}S_{\mathcal{A}t}+\varDelta_tS_{\mathcal{A}t}+\frac{1}{2}(S_{\mathcal{A}t},S_{\mathcal{A}t})_t=0
\label{centext1}
\end{equation}
(cf. eq. \ceqref{bvfix7}), $d_{\mathcal{A}}$ being the differential of $\mathcal{A}$, and the RGE \pagebreak
\begin{equation}
\frac{dS_{\mathcal{A}t}}{dt}=\varphi^\bcdot{}_tS_{\mathcal{A}t}+r^\bcdot{}_{\varphi t}, 
\label{centext2}
\end{equation}
(cf. eq. \ceqref{bvfix19}) also modulo $\mathcal{A}$--valued constants. 
Eq. \ceqref{centext1} arises from demanding that the derived BV effective covariant Laplacian
\begin{equation}
\varDelta_{tS\mathcal{A}}f=d_{\mathcal{A}}f+\varDelta_tf+(S_{t\mathcal{A}},f)_t. 
\label{centext3}
\end{equation}
(cf. eq. \ceqref{bvfix9}) is nilpotent on $\mathcal{A}$--valued functionals $f$ on $T^*[-1]\mathcal{F}$. 
The reason why  $S_{\mathcal{A}t}$ is defined and \ceqref{centext1}, \ceqref{centext2} hold 
only up to $\mathcal{A}$--valued constants is that the nilpotence of $\varDelta_{tS\mathcal{A}}$
entails only that $\ad_t(d_{\mathcal{A}}S_{\mathcal{A}t}+\varDelta_tS_{\mathcal{A}t}
+(S_{\mathcal{A}t},S_{\mathcal{A}t})_t/2)=0$. 


The basic problem of derived BV RG theory is finding out under which conditions 
the derived BV EA $S_{\mathcal{A}t}$ can be lifted to an $\mathcal{A}$--valued functional 
on $T^*[-1]\mathcal{F}$ satisfying 
eqs. \ceqref{centext1}, \ceqref{centext2} exactly and not simply modulo
$\mathcal{A}$--valued constants. It is always possible to choose a lift $S_{\mathcal{A}t}$ obeying 
\ceqref{centext2} exactly by choosing a lift $S_{\mathcal{A}s}$ for a certain value $s$ of the RG scale
and defining $S_{\mathcal{A}t}$ for an arbitrary value $t$ of the scale 
by demanding that it obeys the integrated RG flow equation \ceqref{bvfix18}
\begin{equation}
S_{\mathcal{A}t}=\varphi_{t,s}{}^*S_{\mathcal{A}s}+r_{\varphi t,s}.
\label{centext4}
\end{equation}
Then, eq. \ceqref{centext2} holds exactly by straightforward differentiation of \ceqref{centext4}
with respect to $t$. Once this is done, however, eq. \ceqref{centext1} gets weakened as 
\begin{equation}
d_{\mathcal{A}}S_{\mathcal{A}t}+\varDelta_tS_{\mathcal{A}t}+\frac{1}{2}(S_{\mathcal{A}t},S_{\mathcal{A}t})_t=\alpha_{\mathcal{A}},
\label{centext5}
\end{equation}
where $\alpha_{\mathcal{A}}$ is a degree $1$ element of $\mathcal{A}$. It can be verified 
that $\alpha_{\mathcal{A}}$ is independent from $t$, that $d_{\mathcal{A}}\alpha_{\mathcal{A}}=0$ and 
that a different choice of the lift $S_{\mathcal{A}t}$ alters $\alpha_{\mathcal{A}}$ by an amount 
$d_{\mathcal{A}}\beta_{\mathcal{A}}$ for some degree $0$ element $\beta_{\mathcal{A}}$ of $\mathcal{A}$.
Therefore, there exists a degree $1$ cohomology class $[\alpha_{\mathcal{A}}]$ in $H^1(\mathcal{A},d_{\mathcal{A}})$
independent from the scale $t$ that obstructs the existence of a lift $S_{\mathcal{A}t}$ obeying 
\ceqref{centext1} exactly. 

A standard application of the above derived geometric set--up occurs in the analysis of BV EA homotopies
\ccite{Costello:2016faqft}. In such a case, $\mathcal{A}=\Omega^*(I)$ is the algebra of differential forms in the 
interval $I=[0,1]$ and $d_{\mathcal{A}}=d_{dR}$ is the Rham differential. The derived BV EA is of the
form $S_{\Omega^*(I)t}=S_{xt}+d_{dR}x\,S^*{}_{xt}$, where $S_{xt}$, $S^*{}_{xt}$ have respectively degrees $0$, $-1$.
The obstruction form reads as $\alpha_{\Omega^*(I)}=\alpha_x+d_{dR}x\,\alpha^*{}_x$ with components 
$\alpha_x$, $\alpha^*{}_x$ of degrees $1$, $0$. The deformation $S_{\Omega^*(I)t}$ is liftable provided $\alpha_x=0$. 
In such a case, one has 
\begin{align}
&\varDelta_tS_{xt}+\frac{1}{2}(S_{xt},S_{xt})_t=0,
\vphantom{\Big]}
\label{}
\\
&\frac{\partial S_{xt}}{\partial x}-\varDelta_tS^*{}_{xt}-(S_{xt},S^*{}_{xt})_t=\alpha^*{}_x.
\vphantom{\Big]}
\label{}
\end{align}  
We can readily absorb $\alpha^*{}_x$ by shifting $S_{xt}$ by the irrelevant constant $\int_0^x dx\,\alpha^*{}_x$.
Upon doing so, for fixed $x$, $S_{xt}$ is a BV EA and $\partial S_{xt}/\partial x$ is trivial 
in the $\varDelta_{tS_x}$ cohomology. 

\vspace{.175mm}
In many instances, one considers the case where $\mathcal{A}$ is the Chevalley--Eilenberg algebra
$\CE(\mathfrak{g})$ of an $L_\infty$ algebra $\mathfrak{g}$ and $d_{\mathcal{A}}$ is the associated 
Chevalley--Eilenberg differential $d_{\CE(\mathfrak{g})}$. In such cases, $\mathfrak{g}$ 
encodes a generalized symmetry of the EFT and the derived BV EA $S_{\CE(\mathfrak{g})t}$
describes an action of $\mathfrak{g}$ on the a BV EFT in the interpretation
of ref. \ccite{Costello:2016faqft}. 

\vspace{.175mm}
If $S_{\CE(\mathfrak{g})t}$ can be lifted, then the action of $\mathfrak{g}$ is called inner. As we have just shown, 
the liftability of $S_{\CE(\mathfrak{g})t}$ is equivalent to the vanishing of the obstruction class
$[\alpha_{\CE(\mathfrak{g})}]$ in $H^1(\CE(\mathfrak{g}),d_{\CE(\mathfrak{g})})$.
The cohomology $H^*(\CE(\mathfrak{g}),d_{\CE(\mathfrak{g})})$ is just the Chevalley--Eilenberg cohomology of 
$\mathfrak{g}$. A class $[\alpha_{\CE(\mathfrak{g})}]$ in $H^1(\CE(\mathfrak{g}),d_{\CE(\mathfrak{g})})$ so describes a 
$-1$ shifted central extension of $\mathfrak{g}$.
\begin{equation}
0\longrightarrow \mathbb{C}[-1]\longrightarrow\widehat{\mathfrak{g}}\longrightarrow\mathfrak{g}\longrightarrow 0. 
\label{centext6}
\end{equation}
The non triviality of the extension is thus tantamount to the non liftability of $S_{\CE(\mathfrak{g})t}$.
This is how BV anomalies appear.  

\vspace{1.5mm}

\subsection{\textcolor{blue}{\sffamily Perturbative BV RG theory}}\label{subsec:pertbvrg}

In perturbation theory, one assumes that it is possible to turn off the interactions
of a given quantum field theory and that a free quantum field theory that is fully under 
control and simple is yielded in this way. 

\vspace{.175mm}
In BV theory, the effective free field theory is based on a BV structure characterized by a total 
field space $T^*[-1]\mathcal{F}^0$ and an effective BV form $\omega^0{}_t$ and measure $\mu^0{}_t$ 
and is governed by a free BV EA $S^0{}_t$ all depending on the energy scale 
$t$. To implement the perturbative program, one considers the 
algebra of formal power series $\Fun(T^*[-1]\mathcal{F}^0)[[\hbar]]$ over $\Fun(T^*[-1]\mathcal{F}^0)$
instead of $\Fun(T^*[-1]\mathcal{F}^0)$ itself as the appropriate algebra of field functionals, 
the Planck constant $\hbar$ working here as formal expansion parameter,  
and replaces $S^0{}_t$ by $S^0_t/\hbar$ throughout.

As required, the free BV EA $S^0{}_t$ satisfies the ME \ceqref{bvfix7}, 
\begin{equation}
\hbar\varDelta^0{}_tS^0{}_t+\frac{1}{2}(S^0{}_t,S^0{}_t)^0{}_t=0. 
\label{pertbvrg1}
\end{equation}
Since $S^0_t$ is independent from $\hbar$, the two terms in the left hand side of \ceqref{pertbvrg1} 
must separately vanish and so \ceqref{pertbvrg1} effectively breaks up into two independent equations.

The RG flow of the free BV EA $S^0{}_t$ is governed by a free flow map $\varphi^0{}_{t,s}$.
$S^0{}_t$ obeys accordingly the RGE \ceqref{bvfix19}, \hphantom{xxxxxxxxxxxxx}  
\begin{equation}
\frac{dS^0{}_t}{dt}=\varphi^{0\bcdot}{}_t S^0{}_t+\hbar r^{\bcdot}{}_{\varphi^0t}. 
\label{pertbvrg2}
\end{equation}
Unlike the ME \ceqref{pertbvrg1}, \ceqref{pertbvrg2}
does not split into simpler equations, because $\varphi^{0\bcdot}{}_t$ and $r^{0\bcdot}{}_t$
generally depend on $S^0_t/\hbar$, for reasons explained at the end of subsect. \cref{subsec:bvrenorm},
making the two terms of the right hand side $\hbar$ dependent. 

The full BV RG EA $S_t$ decomposes perturbatively as  
\begin{equation}
S_t=S^0{}_t+I_t, 
\label{pertbvrg3}
\end{equation}
where the interaction action $I_t$ is an element of $\Fun(T^*[-1]\mathcal{F}^0)[[\hbar]]$. 

The full EA $S_t$ also satisfies the ME \ceqref{pertbvrg1},
\begin{equation}
\hbar\varDelta^0{}_tS_t+\frac{1}{2}(S_t,S_t)^0{}_t=0, 
\label{pertbvrg4}
\end{equation}
which on account of the free ME \ceqref{pertbvrg1} 
reduces effectively to 
\begin{equation}
\hbar\varDelta^0{}_{tS^0/\hbar}I_t+\frac{1}{2}(I_t,I_t)^0{}_t=0, 
\label{pertbvrg5}
\end{equation}
where the BV Laplacian $\varDelta^0{}_{tS^0/\hbar}$ is defined according to \ceqref{bvfix9}. 
Since $I_t$ is a formal power series in $\hbar$, this equation 
relates the coefficients of the $\hbar$ powers of $I_t$ in a non trivial manner. 

The RG flow of the full BV EA $S_t$ is governed by a \pagebreak 
full flow map $\varphi_{t,s}$. $S_t$ then obeys the RGE \hphantom{xxxxxxxxxxxxxxxx}  
\begin{equation}
\frac{dS_t}{dt}=\varphi^\bcdot{}_t S_t+\hbar r^\bcdot{}_{\varphi t}.
\label{pertbvrg6}
\end{equation}
Using the decomposition \ceqref{pertbvrg3}, this yields a RGE for $I_t$. 
Since $\varphi_{t,s}$ differs in general from its free counterpart $\varphi^0{}_{t,s}$ in a complicated way 
because of the interaction, $\varphi^\bcdot$, $r^\bcdot{}_{\varphi t}$ are not
straightforwardly reducible to $\varphi^{0\bcdot}{}_t$, $r^{0\bcdot}{}_t$.
For this reason, there is no simple expression for the RGE of $I_t$,
which must be deduced on a case by case basis. We shall came back to this point 
in a specific setting later in subsect. \cref{subsec:permod}. 

In ref. \ccite{Costello:2007ei}, a formulation of the perturbative RG in BV theory
is provided. Its power   and many nice features make its paradigmatic. We hope that the 
BV RG theory presented in this paper may pave the way to alternative equally fruitful 
perturbative formulations, satisfying the RGE

\vfil\eject

\section{\textcolor{blue}{\sffamily Batalin--Vilkovisky renormalization group supersymmetry}}\label{sec:rgsusy}

In this section, we shall introduce the main novelty of the present work, a BV RG set--up enjoying 
an RG supersymmetry imposing important constraints on the form of the resulting RGE. 
This will allow us to establish a connection between the BV RGE we obtain and Polchinski's version
of the RGE \ccite{Polchinski:1983gv}. 

To develop our line of thought, we are forced to make a few natural assumptions. So, our results are to an 
extent conjectural. In particular, we have no general proof that the RG supersymmetry is a general property of BV EFT.
In next section, we shall illustrate non trivial models exhibiting it.


\subsection{\textcolor{blue}{\sffamily BV RG supersymmetry}}\label{subsec:rgsusy}

In BV theory, field space is a graded manifold, field functionals form a graded 
commutative function algebra and basic structures such as the BV form, measure, bracket and 
Laplacian are all graded in the appropriate sense. The ubiquitous occurrence of the mathematical 
structures proper of graded geometry in BV theory suggests that it may be worthy to explore 
whether extension of the RG energy scale parameter manifold $\mathbb{R}$ to a more general 
graded manifold is capable of providing new useful ways of analyzing BV RG theory. 

In what follows, we shall demonstrate that enhancing the energy scale space $\mathbb{R}$ to its shifted tangent 
bundle $T[1]\mathbb{R}$ does indeed furnish under certain conditions important structural information
on the BV RG flow and RGE. To avoid confusion, we shall call the RG set--ups based 
on $\mathbb{R}$ and $T[1]\mathbb{R}$ ``basic'' and ``extended'', respectively. 

Switching from the $\mathbb{R}$ to the $T[1]\mathbb{R}$ parameter space involves 
adding to the usual scale parameter $t$ a degree $1$ partner $\theta$. 
In the extended set--up, so, all the constitutive elements of the BV RG framework
get modified by acquiring a $\theta$ dependence, while maintaining all their formal 
properties. Further, on general grounds, because of the nilpotence of $\theta$, 
they can be expressed as non homogeneous linear polynomials of $\theta$. 

In the extended set--up, the BV form $\omega_{t\theta}$ enjoys the expansion \pagebreak 
\begin{equation}
\omega_{t\theta}=\omega_t+\theta\omega^\star{}_t,
\vphantom{\Big]}
\label{nnbveffact-1}
\end{equation}
where $\omega_t$ is the basic set--up form and $\omega^\star{}_t$ is 
a new partner form. The properties of $\omega_{t\theta}$ as BV form entail that 
$\omega^\star{}_t$ has degree $-2$ and is closed. 
The extended set--up BV measure $\mu_{t\theta}$ is instead independent form $\theta$ 
because of its multiplicative nature and thus equal to its basic counterpart, 
\begin{equation}
\mu_{t\theta}=\mu_t.
\vphantom{\Big]}
\label{nnbveffact0}
\end{equation}

By \ceqref{nnbveffact-1}, the extended set--up BV bracket $(-,-)_{t\theta}$ has the structure
\begin{equation}
(f,g)_{t\theta}=(f,g)_t+\theta (-1)^{|f|}(f,g)^\star {}_t 
\vphantom{\Big]}
\label{nnbveffact6}
\end{equation} 
for $f,g\in \Fun(T^*[-1]\mathcal{F})$, where $(-,-)_t$ is the basic set--up bracket and $(-,-)^\star {}_t$ is 
a new partner bilinear bracket. The properties of $(-,-)_{t\theta}$ as a BV bracket imply that $(-,-)^\star {}_t$ 
has degree $0$, is graded symmetric and derivative in both its arguments. 
The graded Jacobi identity, conversely, fails to hold. By \ceqref{nnbveffact-1} again, the BV Laplacian $\varDelta_{t\theta}$ 
decomposes similarly as 
\begin{equation}
\varDelta_{t\theta}=\varDelta_t+\theta \varDelta^\star {}_t, 
\vphantom{\Big]}
\label{nnbveffact5}
\end{equation}
$\varDelta_t$ being the basic set--up Laplacian and $\varDelta^\star {}_t$ a partner Laplacian,
which by the BV Laplacian properties of $\varDelta_{t\theta}$ has a degree $0$ and commutes with $\varDelta_t$. 

In the extended BV RG framework, 
the BV RG EA $S_{t\theta}$ must likewise be a function of $t$, $\theta$ and therefore
expressible as \hphantom{xxxxxxxxx}
\begin{equation}
S_{t\theta}=S_t+\theta S^\star{}_t, 
\vphantom{\Big]}
\label{nnbveffact6/1}
\end{equation}
where $S_t$ is the basic set--up EA and $S^\star {}_t$ is a degree $-1$ partner of it. 
$S_{t\theta}$ must obey the appropriate extended version of the BV quantum ME \ceqref{bvfix7}, 
\begin{equation}
\varDelta_{t\theta}S_{t\theta}+\frac{1}{2}(S_{t\theta},S_{t\theta})_{t\theta}=0. 
\label{bveffact25/1}
\end{equation}
Insertion of the expansions \ceqref{nnbveffact6}, \ceqref{nnbveffact5} and \ceqref{nnbveffact6/1}
into \ceqref{bveffact25/1} yields two equations. The first is the basic BV quantum ME 
\ceqref{bvfix7}. The second is novel. It relates $S_t$ and $S^\star{}_t$ and reads \pagebreak 
\begin{equation}
\varDelta_tS^\star{}_t+(S_t,S^\star{}_t)_t
=\varDelta^\star{}_tS_t+\frac{1}{2}(S_t,S_t)^\star{}_t.
\label{bveffact26}
\end{equation}
As we shall see in due course, \ceqref{bveffact26} will play a role in shaping the form 
of the extended set--up BV RGE. 

In the extended BV RG framework, the  BV RG flow $\varphi_{t\theta,s\zeta}$ depends on $t,s$, $\theta,\zeta$. 
The obvious generalization of relations \ceqref{bvfix11}--\ceqref{bvfix13} obeyed by $\varphi_{t\theta,s\zeta}$ 
entails that this latter factorizes as 
\begin{equation}
\varphi_{t\theta,s\zeta}=(\id_{T^*[-1]\mathcal{F}}-\zeta\varphi^\star{}_s)\circ \varphi_{t,s}
\circ (\id_{T^*[-1]\mathcal{F}}+\theta\varphi^\star{}_t),
\vphantom{\Big]}
\label{nnbveffact7}
\end{equation}
where $\varphi_{t,s}$ is the basic set--up RG flow and $\varphi^\star{}_t$ is a degree $-1$ vector field 
on $T^*[-1]\mathcal{F}$. The associated logarithmic Jacobian $r_{\varphi t\theta,s\zeta}$ 
similarly read as 
\begin{equation}
r_{\varphi t\theta,s\zeta}=(\id_{\Fun(T^*[-1]\mathcal{F})}+\theta\varphi^\star{}_t)^*
(r_{\varphi t,s}+\theta r^\star{}_{\varphi t}-\zeta\varphi_{t,s}{}^*r^\star{}_{\varphi s}), 
\label{nnbveffact7/1}
\end{equation}
where  $r_{\varphi t,s}$ is  the basic set--up logarithmic Jacobian and $r^\star{}_{\varphi t}$ is a degree 
$-1$ element of $\Fun(T^*[-1]\mathcal{F})$. The canonical nature of the  flow maps $\varphi_{t\theta, s\zeta}$ 
entails that $\varDelta_t$, $\varDelta^\star {}_t$, $\varphi_{t,s}$, $\varphi^\star{}_t$, $r_{\varphi t,s}$ and 
$r^\star{}_{\varphi t}$ obey a host 
of identities which follow from substituting expressions \ceqref{nnbveffact5}, \ceqref{nnbveffact7} and 
\ceqref{nnbveffact7/1} into the identity $\varDelta_{t\theta}\varphi_{t\theta,s\zeta}
-\varphi_{t\theta,s\zeta}\varDelta_{s\zeta}+\ad_{t\theta}r_{\varphi t\theta,s\zeta}\,\varphi_{t\theta,s\zeta}=0$ 
(cf. eq. \ceqref{bvint13}). 

The RG flow of the extended BV EA $S_{t\theta}$ is driven by the flow maps $\varphi_{t\theta,s\zeta}$,
\begin{equation}
S_{t\theta}=\varphi_{t\theta,s\zeta}{}^*S_{s\zeta}+r_{\varphi t\theta,s\zeta}.
\label{nnbveffact7/2}
\end{equation}
At the infinitesimal level,  \ceqref{nnbveffact7/2} reproduces the basic set--up BV RGE \ceqref{bvfix19}
obeyed by $S_t$, but it also yields another relation, namely 
\begin{equation}
S^\star{}_t=\varphi^\star{}_tS_t+r^\star{}_{\varphi t}, 
\label{bveffact16}
\end{equation}
about which we shall have more to say momentarily. 

The degree $-1$ partner $S^\star {}_t$ of the EA $S_t$ allows 
us to write the infinitesimal generator and logarithmic Jacobian 
$\varphi^\bcdot{}_t$ and $r^\bcdot{}_t$ of the basic set--up BV RG flow $\varphi_{t,s}$, 
in terms of which the RGE is written, in the reduced form 
\begin{align}
&\varphi^\bcdot{}_t=-\ad_tS^\star {}_t+\bar\varphi^\bcdot{}_t, 
\vphantom{\Big]}
\label{bveffact18}
\end{align}  
\vspace{-1cm}\eject\noindent
\begin{align}
&r^\bcdot{}_{\varphi t}=\varDelta_tS^\star {}_t+\bar r^\bcdot{}_{\varphi t},
\vphantom{\Big]}
\label{bveffact19}
\end{align}  
$\bar\varphi^\bcdot{}_t$ and $\bar r^\bcdot{}_t$ being a degree $0$ vector field and a degree $0$ 
element of $\Fun(T^*[-1]\mathcal{F})$, called 
reduced infinitesimal generator and Jacobian, respectively. Using these and eq. \ceqref{bveffact26}, 
the basic set--up BV RGE for the EA $S_t$ can be cast as 
\begin{equation}
\frac{dS_t}{dt}=\varDelta^\star{}_tS_t+\frac{1}{2}(S_t,S_t)^\star{}_t
+\bar\varphi^\bcdot{}_tS_t+\bar r^\bcdot{}_{\varphi t}. \vphantom{\bigg]}
\label{bveffact27}
\end{equation}
Recall that, in the RGE \ceqref{bveffact27}, $\varDelta^\star{}_t$ and $(-,-)^\star {}_t$ 
are a degree $0$ second order differential operator and a degree $0$ graded symmetric bracket,
respectively. The first two terms in the right hand side,  
so, have a form analogous to that of the leading terms appearing in Polchinski RGE 
\ccite{Polchinski:1983gv}. Thus, switching to the extended set--up has led us to a
more structured RGE than that one would have in the mere basic RG framework. 
The reduced infinitesimal generator and Jacobian 
$\bar\phi^\bcdot{}_t$ and $\bar r^\bcdot{}_t$ turn out under favorable conditions   
to be simpler than their unreduced counterparts $\phi^\bcdot{}_t$ and $r^\bcdot{}_t$. 
We expect them to correspond to the so--called ``seed'' terms of the RGE, though at the 
present level of generality we cannot show this. 

To understand the structure of the RGE \ceqref{bveffact27}, we recall that the RG flow of the BV EA $S_t$ 
leaves the partition function invariant. Since the BV structure encoded in the scale dependent
BV bracket $(-,-)_t$ and measure $\mu_t$ does itself vary along the flow, the flow rate 
$dS_t/dt$ of $S_t$ should comprise in principle two components. The first is such that it 
would leave the partition function invariant if the variation of the BV structure 
did not occur. This should therefore be a 
degree $0$ $\varDelta_{tS}$ coboundary. The second compensates for the actual variation 
of the partition function caused by the first due to the variation of the BV structure. 
This cannot be a degree $0$ $\varDelta_{tS}$ coboundary and should not be even a cocycle. 

Eq. \ceqref{bveffact26} shows that the sum of the first two terms in the right hand side of the RGE 
\ceqref{bveffact27} is a degree $0$ $\varDelta_{tS}$ coboundary. 
This contribution is so naturally identified with the first component of the flow rate 
$dS_t/dt$ hypothesized in the previous paragraph. 
The sum of the last two terms in the right hand side of \ceqref{bveffact27}
should thus constitute the second component. Note that the degree $-1$ partner $S^\star{}_t$
of $S_t$ acts as the $\varDelta_{tS}$ integral of the first component.
Beyond this, its role  needs to be further elucidated.
More about this in subsect. \cref{subsec:pertrgsusy}. 

\vspace{-1.5mm}\eject

The results which we have obtained hinge on the extension of the energy scale space from $\mathbb{R}$ 
to $T[1]\mathbb{R}$ by adjoining to the usual scale parameter $t$ an odd partner $\theta$, leading effectively 
to a sort of RG supersymmetry. Whether there is something more profound behind this seemingly formal 
procedure we cannot say yet. 

Before concluding this subsection, it must be emphasized that in the above analysis 
we have assumed throughout that the enhancement of the basic set--up
to the extended one is possible. However, this cannot be shown in general but only on a case by case basis. 
In sect. \cref{sec:models}, we shall construct a non trivial model admitting an extension.

\subsection{\textcolor{blue}{\sffamily BV RG flow as a BV MA homotopy}}\label{subsec:bvhomotop}

Identifying the degree $1$ coordinate $\theta$ of $T[1]\mathbb{R}$ of the extended RG set--up 
with \linebreak the de Rham differential $d_{dR}t$, the BV EA $S_{t\theta}$ given in 
\ceqref{nnbveffact6/1} can be written as $S_t+d_{dR}t S_t{}^*$ and so be regarded as a functional on $T^*[-1]\mathcal{F}$ 
valued in the differential graded commutative algebra $\Omega^*(\mathbb{R})$. As such, it represents
a homoto\-py relating the BV MAs at different values of the RG scale $t$
\footnote{We thank the paper's referee for pointing this out to us.}. This homotopy however cannot be described
in the derived geometric framework of subsect. \cref{subsec:centext}, since in that set--up homotopies relate 
BV MAs at a fixed value of $t$. For homotopies of that type, the rate of variation 
$\partial S_{xt}/\partial x$ is always trivial in the degree $0$ $\varDelta_{tS_x}$ cohomology, while,
as discussed at the end of subsect. \cref{subsec:rgsusy}, the flow derivative $dS_t/dt$ is not and cannot be. 

The possibility of viewing the RG flow as a homotopy of BV MAs is implicit in Costello's work
\ccite{Costello:2007ei}, 
but to the best of our knowledge its implications have never been explored to the extent 
we have done in the present paper. 

\subsection{\textcolor{blue}{\sffamily Perturbative BV RG theory and BV RG supersymmetry}}\label{subsec:pertrgsusy}

In this subsection, we shall explore the implications of the RG supersymmetry 
discovered in subsect. \cref{subsec:rgsusy} for perturbative BV RG theory.

When we try to formulate perturbative BV RG theory in the extended set--up, 
we have to assume that the free field theory is characterized by an extended 
BV form $\omega^0{}_{t\theta}$ and measure $\mu^0{}_{t\theta}$ and the associated 
bracket $(-,-)^0{}_{t\theta}$ and Laplacian $\varDelta^0{}_{t\theta}$ 
with the properties described in subsect. \cref{subsec:rgsusy}. 

Besides the above free BV structure, we have also to assume the existence 
of an extended free BV RG EA action $S^0{}_{t\theta}$ so that the basic free EA $S^0{}_t$ is adjoined by 
a free degree $-1$ partner $S^{0\star}{}_t$. Further, not only $S^0{}_t$ satisfies the 
BV quantum ME \ceqref{pertbvrg1}, but $S^0{}_t$ and $S^{0\star}{}_t$ together obey the ME \ceqref{bveffact26},
here reading as 
\begin{equation}
\hbar\varDelta^0{}_tS^{0\star}{}_t+(S^0{}_t,S^{0\star}{}_t)^0{}_t
=\hbar\varDelta^{0\star}{}_tS^0{}_t+\frac{1}{2}(S^0{}_t,S^0{}_t)^{0\star}{}_t.  
\label{ssbveffact1}
\end{equation}
As eq.  \ceqref{pertbvrg1}, this too breaks up effectively into two equations 
due to the $\hbar$ independence of $S^0{}_t$, $S^{0\star}{}_t$. 

A free BV RG flow $\varphi^0{}_{t\theta,s\zeta}$
governs the $t,\theta$ dependence of the extended free EA $S^0{}_{t\theta}$. 
Infinitesimally, this reproduces the basic free BV RGE \ceqref{pertbvrg2} for $S^0{}_t$
and further relates $S^0{}_t$ and $S^{0\star}{}_t$ as in \ceqref{bveffact16}
\begin{equation}
S^{0\star}{}_t=\varphi^{0\star}{}_tS^0{}_t+\hbar r^{\star}{}_{\varphi^0 t}. 
\label{ssbveffact9}
\end{equation} 
By virtue of \ceqref{ssbveffact1}, further, the RGE \ceqref{pertbvrg2},
can be written in the Polchinski form \ceqref{bveffact27}, 
\begin{equation}
\frac{dS^0{}_t}{dt}=\hbar\varDelta^{0\star}{}_tS^0{}_t+\frac{1}{2}(S^0{}_t,S^0{}_t)^{0\star}{}_t
+\bar\varphi^{0\bcdot}{}_tS^0{}_t+\hbar\bar r^\bcdot{}_{\varphi^0 t}.
\label{ssbveffact10}
\end{equation} 

In the extended perturbative BV RG framework, the full BV RG EA $S_{t\theta}$ decomposes 
in analogy to \ceqref{pertbvrg3} as 
\begin{equation}
S_{t\theta}=S^0{}_{t\theta}+I_{t\theta}, 
\label{ssbveffact2}
\end{equation}
where the extended interaction action $I_{t\theta}$ belongs to the formal power series algebra 
$\Fun(T^*[-1]\mathcal{F}^0)[[\hbar]]$. This latter enjoys in turn the expansion 
\begin{equation}
I_{t\theta}=I_t+\theta I^\star{}_t, 
\label{ssbveffact3}
\end{equation}
where $I_t$ is the basic set--up interaction action and $I^\star{}_t$ is a degree 
$-1$ partner of it. In this way, in addition to the perturbative splitting \ceqref{pertbvrg3}, we have also
\begin{equation}
S^\star{}_t=S^{0\star}{}_t+I^\star{}_t. 
\label{ssbveffact4}
\end{equation}
The full EA $S_{t\theta}$ obeys further the extended quantum ME \ceqref{bveffact25/1}
(with $(-,-)_{t\theta}$ and $\varDelta_{t\theta}$ replaced by $(-,-)^0{}_{t\theta}$ and $\hbar\varDelta^0{}_{t\theta}$). 
Insertion of the perturbative expansion \ceqref{ssbveffact2} into this latter 
reproduces upon taking eqs. \ceqref{pertbvrg1}, \ceqref{ssbveffact1} into account
eq. \ceqref{pertbvrg5} and a further equation 
\begin{equation}
\hbar\varDelta^0{}_{tS^0/\hbar}I^\star{}_t+(I_t,I^\star{}_t)^0{}_t
=\hbar\varDelta^{0\star}{}_{tS^0/\hbar}I_t+\frac{1}{2}(I_t,I_t)^{0\star}{}_t,
\label{ssbveffact5}
\end{equation}
where the BV Laplacian $\varDelta^0{}_{tS^0/\hbar}$ is defined according to \ceqref{bvfix9}
and $\varDelta^{0\star}{}_{tS^0/\hbar}$ is the degree $0$ Laplacian given by 
\begin{equation}
\varDelta^{0\star}{}_{tS^0/\hbar}f
=\varDelta^{0\star}{}_tf+\hbar^{-1}(S^{0\star}{}_t,f)^0{}_t+\hbar^{-1}(S^0{}_t,f)^{0\star}{}_t
\label{ssbveffact6}
\end{equation}
for $f\in\Fun(T^*[-1]\mathcal{F}^0)$.  

As in the basic perturbative  RG framework of subsect. \cref{subsec:pertbvrg}, the 
extended BV RG flow $\varphi_{t\theta,s\zeta}$ that 
governs the $t,\theta$ dependence of the extended full EA $S_{t\theta}$ 
differs in principle from it free counterpart $\varphi^0{}_{t\theta,s\zeta}$ because of the effect 
of the interactions.
Therefore, the RGE equation for $S_{t\theta}$ is not simply that for $S^0{}_{t\theta}$ 
with $S^0{}_{t\theta}$ replaced by $S_{t\theta}$. In particular, the basic full EA $S_t$ obeys 
the RGE \ceqref{pertbvrg6}, the $S^\star{}_t$ to $S_t$ relation 
\ceqref{bveffact16} takes the form 
\begin{equation}
S^\star {}_t=\varphi^\star{}_tS_t+\hbar r^\star{}_{\varphi t}
\label{ssbveffact11}
\end{equation}
and the Polchinski RGE \ceqref{bveffact27} reads 
\begin{equation}
\frac{dS_t}{dt}=\hbar\varDelta^{0\star}{}_tS_t+\frac{1}{2}(S_t,S_t)^{0\star}{}_t
+\bar\varphi^\bcdot{}_tS_t+\hbar\bar r^\bcdot{}_{\varphi t}.
\label{ssbveffact12}
\end{equation}
However, as we already mentioned in subsect \cref{subsec:rgsusy}, we expect that the reduced full  
infinitesimal generator and logarithmic Jacobian $\bar\varphi^\bcdot{}_t$ and $\bar r^\bcdot{}_{\varphi t}$
to be somewhat related to the seed terms of the RGE, which, in a perturbative framework, are normally 
determined by the free EA $S^0{}_t$ only and are independent from interactions. For this reason, we {\it assume} that 
in the RGE \ceqref{ssbveffact12}, one has 
\begin{align}
&\bar\varphi^\bcdot{}_t=\bar\varphi^{0\bcdot}{}_t,
\vphantom{\Big]}
\label{ssbveffact13}
\\
&\bar r^\bcdot{}_{\varphi t}=\bar r^\bcdot{}_{\varphi^0 t}.
\vphantom{\Big]}
\label{ssbveffact14}
\end{align}  
Taking this for granted, \pagebreak we can now subtract the free and full RGE \ceqref{ssbveffact10} and \ceqref{ssbveffact12} 
and obtain an RGE equation for the interaction action $I_t$. Upon using relation \ceqref{bveffact18} 
for $\varphi^0{}_{t,s}$ and $S^0{}_t$, we find
\begin{equation}
\frac{dI_t}{dt}=\hbar\varDelta^{0\star}{}_tI_t+\frac{1}{2}(I_t,I_t)^{0\star}{}_t
+(S^0{}_t,I_t)^{0\star}{}_t+\bar\varphi^{0\bcdot}{}_tI_t. 
\label{ssbveffact15}
\end{equation}

Combining relations \ceqref{bveffact18},  \ceqref{bveffact19} for the free and full case with the perturbative decomposition 
\ceqref{ssbveffact4} and taking further identities \ceqref{ssbveffact13}, \ceqref{ssbveffact14}
into account, the following revealing equations are found
\begin{align}
&\varphi^\bcdot{}_t=
\varphi^{0\bcdot}{}_t-\ad^0{}_tI^\star {}_t,
\vphantom{\Big]}
\label{ssbveffact7}
\\
&r^\bcdot{}_{\varphi t}=
r^\bcdot{}_{\varphi^0 t}+\varDelta^0{}_tI^\star {}_t.
\vphantom{\Big]}
\label{ssbveffact8}
\end{align}  
\ceqref{ssbveffact7}, \ceqref{ssbveffact7} provide an interpretation of interaction action partner 
$I^\star{}_t$: it drives the deviation of the full BV RG flow $\varphi_{t,s}$ from its free
counterpart $\varphi^0{}_{t,s}$. 

Analogously to its non perturbative counterpart, eq. \ceqref{bveffact26},
discussed in subsect. \cref{subsec:rgsusy}, eq. \ceqref{ssbveffact5} expresses roughly 
the triviality in the degree $0$ $\varDelta_{tS_/\hbar}$ cohomology of the non seed terms of the RGE \ceqref{ssbveffact15}, 
the degree $-1$ partner $I^\star{}_t$ of the interaction action $I_t$ being their integral. On account of \ceqref{pertbvrg3} 
and \ceqref{ssbveffact2}, eq. \ceqref{ssbveffact11} provides implicitly $I^\star{}_t$ in terms of $I_t$ and so works 
as an integration formula. We have not been able to recast it in a way that neatly separates the free and interacting 
contributions. More work is required to elucidate this point. 

The conclusions of our analysis rest of course on the assumptions \ceqref{ssbveffact13}, \ceqref{ssbveffact14}
and thus fail if these do not obtain. Again, one must check them on a case by case basis. 
In the model detailed in sec. \cref{sec:models}, they are verified to hold. 
It is also found that the last two seed-like terms in the right hand side of \ceqref{ssbveffact15} cancel out
leaving a seedless RGE.


\vfil\eject

\vfil\eject

\section{\textcolor{blue}{\sffamily Models of  Batalin--Vilkovisky  renormalization group}}\label{sec:models}

An important issue in the BV theory of the RG developed in this paper is the construction 
of non trivial models which exemplify it. A program of this scope  
certainly cannot be carried out to its full extent in the limited space of this paper. Still, a few 
simple but non trivial models can be built. 

In what follows, we work in the degree $-1$ symplectic framework originally developed 
by Costello in ref. \ccite{Costello:2007ei}, which has a very rich structure and lends itself 
particularly well to our task. We first review the framework to set our notation.
We then illustrate a free model of BV RG flow and EA both in the basic and  extended set--ups of subsect. 
\cref{subsec:rgsusy}. Finally, we explore the implications of results found for perturbation theory. 

All the statements made below hold strictly speaking in finite dimension.  
Presumably, they can extended also to an infinite dimensional context with limited modifications.


\subsection{\textcolor{blue}{\sffamily The degree --1 symplectic set--up}}\label{subsec:sumbvrg}

In this technical subsection, we review the degree $-1$ symplectic set--up 
of ref. \ccite{Costello:2007ei}, which we shall employ to construct a class of 
models of BV RG theory. A few original results are also presented along the way. 
A more thorough account of these matter will be provided in ref. \ccite{Zucchini:2017ip}. 

We begin by setting our notation.
The basic algebraic structures we shall be concerned with are a graded vector space $\mathcal{E}$,
its dual vector space $\mathcal{E}{}^*$ and its internal endomorphism algebra $\End(\mathcal{E})$. 
We shall consider further the full degree prolongations $E=\mathcal{E}\otimes G_{\mathbb{R}}$, $E^*=\mathcal{E}^*\otimes G_{\mathbb{R}}$
and $\End(E)=\End(\mathcal{E})\otimes G_{\mathbb{R}}$ of $\mathcal{E}$, $\mathcal{E}^*$ and $\End(\mathcal{E})$, where 
$G_{\mathbb{R}}=\bigoplus_{p\in\mathbb{Z}}\mathbb{R}[p]$. $E$ is a graded vector space, $E^*$ is its dual vector space
and $\End(E)$ is its internal endomorphism algebra, as suggested by the notation. For each $p\in\mathbb{Z}$, 
$E_p$, $E^*{}_p$ and $\End_p(E)$ are just respectively $\mathcal{E}$, $\mathcal{E}^*$ and $\End(\mathcal{E})$ 
with degree reset to the uniform value $p$. We shall denote by $|\hspace{-1.3pt}-\hspace{-1.3pt}|$ the degree map of all the graded 
spaces and algebras considered above. 
We shall work mostly with $E$, $E^*$, $\End(E)$. 

\vspace{.175mm}
Homogeneous bases of $\mathcal{E}$, $\mathcal{E}^*$ come in dual pairs $a_i$, $a^{*i}$ with the property that
$|a_i|+|a^{*i}|=0$. Below, we shall set $\epsilon^i=-|a_i|=|a^{*i}|$ for convenience. Given a dual basis pair $a_i$, $a^{*i}$, 
we can expand homogeneous vectors $e\in E$, covectors $l\in E^*$ and endomorphisms $A\in\End(E)$ as $e=a_ie^i$, $l=l_ia^{*i}$ and 
$A=a_i\otimes A^i{}_ja^{*j}$, where $|e^i|=|e|+\epsilon^i$, $|l_i|=|l|-\epsilon^i$ and $|A^i{}_j|=|A|+\epsilon^i-\epsilon^j$. 

\vspace{.175mm}
A homogeneous endomorphism $A\in\End(E)$ is characterized by its graded trace, which is defined as \hphantom{xxxxxxxxxx}
\begin{equation}
\grtr(A)=(-1)^{(|A|+1)\epsilon^i}A^i{}_i. 
\label{sumbvrg-5}
\end{equation}
It can be shown that $\grtr(A)$ is independent from basis choices. Further, one has $|\grtr(A)|=|A|$ and 
$\grtr(AB)=(-1)^{|A||B|}\grtr(BA)$ for $A,B\in \End(E)$. See ref. \ccite{Covolo:2012gt} for an analogous 
notion. 

\vspace{.175mm}
In the symplectic set--up, the graded vector space $\mathcal{E}$ is equipped with a degree $-1$ symplectic 
pairing, i. e.  an antisymmetric bilinear form 
$\langle-,-\rangle_{\mathcal{E}}:\mathcal{E}\times\mathcal{E}\rightarrow\mathbb{R}$ 
with the property that 
for homogeneous $e,f\in \mathcal{E}$, $\langle e,f\rangle_{\mathcal{E}}=0$ whenever 
$|e|+|f|\not=1$.  $\langle-,-\rangle_{\mathcal{E}}$ induces a bilinear pairing 
$\langle-,-\rangle_E:E\times E\rightarrow G_{\mathbb{R}}$ such that 
$\langle-,-\rangle_E:E_p\times E_q\rightarrow \mathbb{R}[-p-q+1]$ and, 
in particular,  $\langle-,-\rangle_E:E_0\times E_0$ $\rightarrow \mathbb{R}[1]$. 

\vspace{.175mm}
The pairing $\langle-,-\rangle_{\mathcal{E}}$ can be viewed as degree $-1$ vector space isomorphism 
$\varpi_{\mathcal{E}}:\mathcal{E}\rightarrow\mathcal{E}^*$. With respect to a given basis $a^{*i}$ of 
$\mathcal{E}^*$, $\varpi_{\mathcal{E}}$ can be expanded as $\varpi_{\mathcal{E}}=a^{*i}\otimes\omega_{ij}a^{*i}$, 
where $\omega_{ij}$ is a non singular antisymmetric real matrix 
such that $\omega_{ij}=0$ for $\epsilon^i+\epsilon^j+1\not=0$. 

\vspace{.175mm}
Using the matrix $\omega_{ij}$ and its inverse $\omega^{ij}$, one can define the symplectic dual bases $a^i$ and $a^*{}_i$ 
of the bases $a_i$ and $a^{*i}$  of an (algebraically) dual pair  by $a^i=a_j\omega^{ji}$ 
and $a^*{}_i=-\omega_{ij}a^{*j}$. These have degrees $|a^i|=-\epsilon_i$ and $|a^*{}_i|=\epsilon_i$, respectively,
where $\epsilon_i=-\epsilon^i-1$. Vectors $e\in E$, covectors $l\in E^*$ and endomorphisms $A\in\End(E)$ 
can be expanded with respect the dual bases $a^i$, $a^*{}_i$ e. g.
$e=-a^ie_i$, $l=-l^ia^*{}_i$, $A=a^i\otimes A_i{}^j a^*{}_j=-a_i\otimes A^i{}^j a^*{}_j=-a^i\otimes A_i{}_j a^*{}^j$
etc. The components with respect the dual bases are related to those with respect to the given bases as expected,
e. g. $e_i=-\omega_{ij}e^j$, $l^i=l_j\omega^{ji}$, $A_i{}^j=-\omega_{ik}A^k{}_l\omega^{lj}$, etc. Care must be taken
when dealing with signs. 

\vspace{.175mm}
The transpose of a homogeneous endomorphism $A\in\End(E)$ is the homogeneous endomorphism 
$A^\sim\in\End(E)$ such that $\langle e,Af\rangle_E=\langle f,A^\sim e\rangle_E$
for $e,f\in E_0$. The matrix components of $A^\sim$ are given in terms of those of $A$ 
by the expression
\begin{equation}
A^\sim{}_i{}^j=(-1)^{|A|(\epsilon^i+\epsilon^j+1)+\epsilon^i(\epsilon^j+1)}A^j{}_i. 
\label{sumbvrg-1}
\end{equation}
One has $|A^\sim|=|A|$ and $(AB)^\sim=(-1)^{1+|A||B|}B^\sim A^\sim$ for $A,B\in\End(E)$ and $1_E{}^\sim=-1_E$.
Furthermore, one has $\grtr A^\sim=\grtr A$, so that $\grtr A=0$ whenever $A^\sim=-A$. 

Henceforth, we concentrate on the degree $0$ subspace $E_0\subset E$. $E_0$ can be endowed with 
a structure of graded manifold through a set of globally defined graded coordinates $x^i$. 
Upon picking a basis $a^{*i}$ of $\mathcal{E}^*$, the $x^i$ are the elements of $E^*$ 
of degree $|x^i|=\epsilon^i$ corresponding to  $a^{*i}$. 
The symplectic dual coordinates $x_i=-\omega_{ij}x^j$ of degree $|x_i|=\epsilon_i$ can also be
used as coordinates of $E_0$. 

$E_0$ is naturally a BV manifold isomorphic to the canonical BV manifold 
$T^*[-1]F_0$ of some graded vector space $F_0$ regarded as a graded manifold.
The Darboux coordinates of $E_0$ are the coordinate functions $x^i$ 
associated to a chosen Darboux basis $a_i$ for the symplectic pairing
$\langle-,-\rangle_{\mathcal{E}}$. The canonical BV form of $E_0$ is the degree $-1$ symplectic form 
\begin{equation}
\omega_E=\frac{1}{2}dx^i\omega_{ij}dx^j.  
\label{sumbvrg3}
\end{equation}
The canonical BV measure is simply \hphantom{xxxxxxxxxxxxxx}
\begin{equation}
\mu_E=\ppp_i dx^i.  
\label{sumbvrg3/1}
\end{equation}


Let $\Fun(E_0)$ be the internal graded commutative algebra of smooth functions on $E_0$
with the usual grading $|\hspace{-1.3pt}-\hspace{-1.3pt}|$. 
The canonical BV bracket $(-,-)_E$ on $\Fun(E_0)$ associated with $\omega_E$ reads 
\begin{equation}
(u,v)_E=(-1)^{\epsilon^i(|u|+1)}\partial_iu\,\omega^{ij}\partial_jv  
\label{nsumbvrg1}
\end{equation}
for $u,v\in\Fun(E_0)$. Further, the canonical BV Laplacian $\varDelta_E$ associated with $\omega_E$ and $\mu_E$ 
is given by \hphantom{xxxxxxxxxxxxxxxxxxxxx}
\begin{equation}
\varDelta_Eu=\frac{1}{2}(-1)^{\epsilon^i}\partial_i\,\omega^{ij}\partial_ju  \pagebreak 
\label{nsumbvrg2}
\end{equation}
for $u\in\Fun(E_0)$. 

The canonical BV structure of $E_0$ specified by the BV form and measure $\omega_E$, $\mu_E$
is not sufficient for the formal developments of later subsections. A suitable generalizations of 
it is required. It is possible to construct a deformation $\omega_A$, $\mu_A$ of $\omega_E$, $\mu_E$ 
for any endomorphism $A\in\End(E)$ that satisfies $|A|=0$ and $A^\sim=-A$ and is invertible.
The $A$--deformed BV form reads 
\begin{equation}
\omega_A=\frac{1}{2}dx^i\omega_{ik}A^{-1k}{}_jdx^j.  
\label{nnsumbvrg3}
\end{equation}
The $A$--deformed BV measure equals instead the canonical one \hphantom{xxxxxxxxxxxxxx}
\begin{equation}
\mu_A=\ppp_i dx^i.
\label{nsumbvrg3/1}
\end{equation}
The $A$--deformed BV bracket on $\Fun(E_0)$ takes in this way the the form 
\begin{equation}
(u,v)_A=(-1)^{\epsilon^i(|u|+1)}\partial_iu\,A^i{}_k\omega^{kj}\partial_jv,
\label{nsumbvrg3}
\end{equation}
while the $A$--deformed BV Laplacian reads 
\begin{equation}
\varDelta_Au=\frac{1}{2}(-1)^{\epsilon^i}\partial_i\,A^i{}_k\omega^{kj}\partial_j u,
\label{nsumbvrg4}
\end{equation}
where again $u,v\in\Fun(E_0)$.  

When $A=1_E$, the $A$--deformed BV form and measure reduce to the canonical ones as do the deformed 
BV bracket and Laplacian. It is convenient however to proceed in another direction 
and relax some of the restriction on the deforming endomorphism $A\in\End(E)$ 
imposing only that $A^\sim=-A$ but requiring that neither 
$|A|=0$ nor $A$ is invertible. The $A$ deformed BV bracket $(-,-)_A$ and Laplacian $\varDelta_A$
can still be defined through \ceqref{nsumbvrg3} and \ceqref{nsumbvrg4}. 
$(-,-)_A$ exhibits properties generalizing those of the canonical 
BV bracket $(-,-)_E$ except for the graded Jacobi identity. Similarly, 
$\varDelta_A$ enjoys properties extending those of the canonical BV Laplacian $\varDelta_E$
except for nilpotence. For $(-,-)_A$ and $\varDelta_A$ to have all the 
properties of a BV bracket and Laplacian, respectively, it is required in addition that $|A|=0$ 
mod $2$. 

In the calculations carried out below, we use repeatedly a host of basic identities, which we collect 
here for convenience in index free form and whose proof will be given in \ccite{Zucchini:2017ip}. 
A part of these involve vector fields of the basic form 
\begin{equation}
\langle x,K\ad_E x\rangle_E,
\label{nsumbvrg9}
\end{equation}
where $K\in\End(E)$  is an endomorphism of $E$ such that $K^\sim=-K$. 
These act naturally on deformed brackets:
for $A,B\in\End(E)$ such that $A^\sim=-A$, $|A|=0$ mod $2$ and $B^\sim$ $=-B$, one has 
\begin{align}
&\langle x,B\ad_E x\rangle_E(u,v)_A=(\langle x,B\ad_E x\rangle_Eu,v)_A
\vphantom{\Big]}
\label{sumbvrg15/1}
\\
&\hspace{2.cm}+(-1)^{|B|(|u|+1)}(u,\langle x,B\ad_E x\rangle_Ev)_A
-(-1)^{|B|(|u|+1)}(u,v)_{AB+BA}
\vphantom{\Big]}
\nonumber
\end{align}
for $u,v\in\Fun(E_0)$. Further, they have simple commutation relations with the associated Laplacians 
\begin{equation}
[\langle x,B\ad_E x\rangle_E,\varDelta_A]=(-1)^{1+|B|}\varDelta_{AB+BA}. 
\label{sumbvrg15}
\end{equation}
Another part concern quadratic functions of $\Fun(E_0)$ of the form 
\begin{equation}
\langle x,Nx\rangle_E
\label{nsumbvrg15}
\end{equation}
with $N\in\End(E)$ an endomorphism of $E$ such that $N^\sim=+N$. 
They are characterized by simple deformed BV brackets: if $A,B,C\in\End(E)$ are endomorphism such that $A^\sim=-A$ 
and $B^\sim=+B$, $C^\sim=+C$, then 
\begin{equation}
(\langle x,Bx\rangle_E,\langle x,Cx\rangle_E)_A=4\langle x,BACx\rangle_E.
\label{sumbvrg25}
\end{equation}
The action of the deformed Laplacian on them is also simple enough,
\begin{equation}
\varDelta_A\langle x,Bx\rangle_E=\grtr(AB). 
\label{sumbvrg24}
\end{equation}
Finally, we have \hphantom{xxxxxxxxxxxxxx}
\begin{equation}
\langle x,A\ad_E x\rangle_E\langle x,Bx\rangle_E
=\langle x,(AB+(-1)^{|A||B|}BA)x\rangle_E.
\label{sumbvrg26}
\end{equation}


\subsection{\textcolor{blue}{\sffamily Symplectic gl(1$|$1) structures}}\label{subsec:n2struct}  

$\mathfrak{gl}(1|1)$ structures, \pagebreak which we review briefly in this subsection, are recurrent in differential geometry, 
supersymmetric quantum mechanics and topological sigma models. An $\mathfrak{gl}(1|1)$ structure enters also  
in the symplectic set--up of ref. \ccite{Costello:2007ei} as one of its constitutive elements.

Let $\mathcal{E}$ be degree $-1$ symplectic vector space (cf. subsect.  \ceqref{subsec:sumbvrg}). 
An $\mathfrak{gl}(1|1)$ structure on $\mathcal{E}$ consists of four endomorphisms
$Q,\overline{Q},H,F\in\End(E)$ of degrees $|Q|=1$, $|\overline{Q}|=-1$, $|H|=0$, $|F|=0$ 
satisfying the graded commutation relations
\begin{align}
&[Q,Q]=0, \qquad [\overline{Q},\overline{Q}]=0,
\vphantom{\Big]}
\label{n2struct4}
\\
&[Q,\overline{Q}]=H,
\vphantom{\Big]}
\label{n2struct5}
\\
&[Q,H]=[\overline{Q},H]=0,
\vphantom{\Big]}
\label{n2struct6}
\\
&[F,Q]=Q,\qquad [F,\overline{Q}]=-\overline{Q},
\vphantom{\Big]}
\label{n2struct8}
\\
&[F,H]=0
\vphantom{\Big]}
\label{n2struct9}
\end{align}
and the transposition conditions
\begin{align}
&Q^\sim=Q,
\vphantom{\Big]}
\label{n2struct1}
\\
&\overline{Q}^\sim=-\overline{Q},
\vphantom{\Big]}
\label{n2struct2}
\\
&H^\sim=-H,
\vphantom{\Big]}
\label{n2struct3}
\\
&F^\sim=F+1_E.
\vphantom{\Big]}
\label{n2struct3/0}
\end{align}
\noindent
A $\mathfrak{gl}(1|1)$ structure on $\mathcal{E}$ is named in this way because the \ceqref{n2struct4}--\ceqref{n2struct9} 
are the basic Lie brackets of the $\mathfrak{gl}(1|1)$ Lie superalgebra. 
$Q$, $\overline{Q}$, $H$ and $F$ are called respectively supercharge, conjugate supercharge 
Hamiltonian and Fermion number in the physical literature. 

Relations \ceqref{n2struct5}, \ceqref{n2struct8} imply that 
\begin{align}
\grtr Q=0,
\vphantom{\Big]}
\label{n2struct7}
\\
\grtr\overline{Q}=0,
\vphantom{\Big]}
\label{n2struct7/1}
\\
\grtr H=0.
\vphantom{\Big]}
\label{n2struct7/2}
\end{align}
$Q,\overline{Q},H$ constitute \pagebreak in this way an $\mathfrak{sl}(1|1)$ structure on $E$, since the 
\ceqref{n2struct4}--\ceqref{n2struct6} are the standard Lie brackets of the $\mathfrak{sl}(1|1)$ 
Lie superalgebra. 

The fact that $|H|=0$ makes it possible to construct an endomorphism 
$\phi(H)\in\End(E)$ for any real analytic function $\phi(x)$. By \ceqref{n2struct6}, \ceqref{n2struct9}
and \ceqref{n2struct3}, $\phi(H)$ has zero degree, $|\phi(H)|=0$, 
commutes with $Q$, $\overline{Q}$ and $F$
\begin{align}
&[Q,\phi(H)]=[\overline{Q},\phi(H)]=0,
\vphantom{\Big]}
\label{n2struct11}
\\
&[F,\phi(H)]=0
\vphantom{\Big]}
\label{n2struct11/0}
\end{align}
and transposes as \hphantom{xxxxxxxxxxxxxxxxxxx}
\begin{equation}
\phi(H)^\sim=-\phi(H). 
\label{n2struct10}
\end{equation}

The property that $|F|=0$ allows likewise the introduction of the sign endomorphism $(-1)^F\in\End(E)$. 
By \ceqref{n2struct8}, \ceqref{n2struct9} and \ceqref{n2struct3/0}, $(-1)^F\in\End(E)$ has degree $0$, $|(-1)^F|=0$,
anticommutes with $Q$, $\overline{Q}$ and commutes with $H$
\begin{align}
&\{Q,(-1)^F\}=\{\overline{Q},(-1)^F\}=0, 
\vphantom{\Big]}
\label{n2struct12}
\\
&[H,(-1)^F]=0,
\vphantom{\Big]}
\label{n2struct13}
\end{align}
where $\{A,B\}=AB+(-1)^{|A||B|}BA$ denotes the graded anticommutator of two endomorphisms $A,B\in \End(E)$, 
and transposes as 
\begin{equation}
(-1)^{F\sim}=+(-1)^F. 
\label{n2struct14}
\end{equation}


\subsection{\textcolor{blue}{\sffamily Free models of BV RG}}\label{subsec:freemod}  

The basic datum required for the construction of free models of BV RG illustrated below 
is a degree $-1$ symplectic vector space $\mathcal{E}$ together with a $\mathfrak{gl}(1|1)$ structure
$Q,\overline{Q},H, F$ on $\mathcal{E}$. 

Our aim is working out  a free BV EFT on the field space $E_0=T^*[-1]F_0$. To this end, as 
explained in subsect. \cref{subsec:bvrenorm}, we have to replace the unregularized BV form and measure 
$\omega_E$, $\mu_E$ with effective counterparts $\omega_t$, $\mu_t$
depending on an energy scale parameter $t$. 
Inspired by Costello 's formulation \ccite{Costello:2007ei}, we choose $\omega_t$, $\mu_t$
to be the deformations of $\omega_E$, $\mu_E$ associated with the endomorphism $\ee^{-tH}\in\End(E)$
in accordance with \ceqref{nnsumbvrg3}, \ceqref{nsumbvrg3/1} 
\begin{align}
&\omega_t=\omega_{\ee^{-tH}},
\vphantom{\Big]}
\label{nnfreemod1}
\\
&\mu_t=\mu_{\ee^{-tH}}=\mu_E.
\vphantom{\Big]}
\label{nnfreemod2}
\end{align}
Since $|\ee^{-tH}|=0$ and $\ee^{-tH\sim}=-\ee^{-tH}$ and $\ee^{-tH}$ is invertible, 
this can be consistently done. 
The effective BV bracket and Laplacian $(-,-)^0{}_t$, $\varDelta^0{}_t$ are then given by 
the concomitant deformations of $(-,-)_E$, $\varDelta_E$, viz
\begin{align}
&(u,v)^0{}_t=(u,v)_{\ee^{-tH}},
\vphantom{\Big]}
\label{freemod7/1}
\\
&\varDelta^0{}_t=\varDelta_{\ee^{-tH}}
\vphantom{\Big]}
\label{freemod1}
\end{align}
with the right hand sides defined  conforming with \ceqref{nsumbvrg3}, \ceqref{nsumbvrg4},
respectively.

The RG flow $\varphi^0{}_{t,s}$ of the free BV EFT is generated infinitesimally by the degree $0$ 
vector field $\langle x,H\ad_E x\rangle_E$, 
\begin{equation}
\varphi^0{}_{t,s}=\exp\bigg(\frac{t-s}{2}\langle x,H\ad_E x\rangle_E\bigg). \vphantom{\bigg]}
\label{freemod2}
\end{equation}
Its logarithmic Jacobian vanishes,  
\begin{equation}
r_{\varphi^0 t,s}=0. 
\label{freemod3}
\end{equation}
The pull--back action $\varphi^0{}_{t,s}{}^*$ of $\varphi^0{}_{t,s}$ on the internal function algebra $\Fun(E_0)$ 
of $E_0$ is given also by the right hand side of \ceqref{freemod2} upon regarding $\langle x,H\ad_E x\rangle_E$
as a derivation of $\Fun(E_0)$.  

The proof that $\varphi^0{}_{t,s}$ is a BV RG flow is simple enough.
To begin with, we note that relations \ceqref{bvfix11}--\ceqref{bvfix13} are trivially satisfied.
To show that the maps $\varphi^0{}_{t,s}$ are canonical, it is enough to prove that   
they intertwine the BV brackets $(-,-)^0{}_t$ as in \ceqref{bvint9}. 
This follows from the relation
\begin{equation}
(u,v)^0{}_t=\varphi^0{}_{t,0}{}^*(\varphi^0{}_{t,0}{}^{-1*}u, \varphi^0{}_{t,0}{}^{-1*}v)_E,
\label{nfreemod1}
\end{equation}
which can be shown straightforwardly using \ceqref{sumbvrg15/1}. 
The vanishing of $r_{\varphi^0 t,s}$ can be inferred by comparing the relation \pagebreak 
\begin{equation}
\varDelta^0{}_t=\varphi^0{}_{t,0}{}^*\varDelta_E\varphi^0{}_{t,0}{}^{-1*},
\label{nfreemod2}
\end{equation}
which is a simple consequence of \ceqref{sumbvrg15}, and \ceqref{bvint13}.

Consider next the one--parameter family of free actions $S^0{}_t\in\Fun(E_0)$,  where 
\begin{equation}
S^0{}_t=-\frac{1}{2}\langle x,Q \ee^{tH}x\rangle_E. 
\label{freemod8}
\end{equation}
We are now going to show that $S^0{}_t$ has the required properties of a BV RG EA.

$S^0{}_t$ obeys the BV quantum ME \ceqref{bvfix7}. This can be immediately using the 
calculations \ceqref{sumbvrg25}, \ceqref{sumbvrg24} and the identities 
\ceqref{n2struct4} and \ceqref{n2struct7},
\begin{equation}
\varDelta^0{}_tS^0{}_t+\,\frac{1}{2}(S^0{}_t,S^0{}_t)^0{}_t
=-\frac{1}{2}\grtr(Q)+\frac{1}{2}\langle x,Q^2\ee^{tH}x\rangle_E=0.
\label{nfreemod3}
\end{equation}

Next,  $S^0{}_t$ satisfies the RG flow relation \ceqref{bvfix18}. Indeed, using the 
expressions \ceqref{freemod2}, \ceqref{freemod8} and relation \ceqref{sumbvrg26}, 
it is not difficult to verify that $\partial(\varphi^0{}_{t,s}{}^*S^0{}_s)/\partial s$ $=0$. 
Therefore,  one has \hphantom{xxxxxxxxxxxxx} 
\begin{equation}
S^0{}_t=\varphi^0{}_{t,t}{}^*S^0{}_t=\varphi^0{}_{t,s}{}^*S^0{}_s
\label{nfreemod4}
\end{equation}
as required. The RGE obeyed by $S^0{}_t$ takes in this way the simple form 
\begin{equation}
\frac{dS^0{}_t}{dt}=\frac{1}{2}\langle x,H(x,S^0{}_t)_E\rangle_E. 
\label{freemod14}
\end{equation}

The free RG set--up considered above is evidently of the basic type discussed in subsect.
\cref{subsec:rgsusy}. It is natural to wonder whether there exists an analogous construction in the 
extended set--up. The answer is affirmative as we show next. 

Our aim is now the construction of an extended version of the basic
free BV EFT on the field space $E_0=T^*[-1]F_0$ worked out above. 
In line with what done earlier in the basic case, we have to replace the 
unregularized BV form and measure $\omega_E$, $\mu_E$ with effective counterparts 
$\omega_{t\theta}$, $\mu_{t\theta}$ depending on an energy scale parameter $t$
and an additional degree $1$ partner parameter $\theta$. We choose $\omega_{t\theta}$, $\mu_{t\theta}$  
to be the deformations of $\omega_E$, $\mu_E$ associated with the endomorphism 
$\ee^{-tH+\theta(-1)^F\overline{Q}}\in\End(E)$ 
according to \ceqref{nnsumbvrg3}, \ceqref{nsumbvrg3/1}, viz
\begin{align}
&\omega_{t\theta}=\omega_{\ee^{-tH+\theta(-1)^F\overline{Q}}}
=\omega_{\ee^{-tH}}+\theta\omega_{\overline{Q}\ee^{-tH}},
\vphantom{\Big]}
\label{nfreemod5}
\end{align}
\vspace{-1cm}\eject\noindent
\begin{align}
&\mu_{t\theta}=\mu_{\ee^{-tH+\theta(-1)^F\overline{Q}}}=\mu_E, \hspace{3cm}
\vphantom{\Big]}
\label{nfreemod6}
\end{align}
where with an abuse of notation we set $\omega_{\overline{Q}\ee^{-tH}}
=\frac{1}{2}\langle dx,\overline{Q}\ee^{tH}dx\rangle_E$. We note that 
$|\ee^{-tH+\theta(-1)^F\overline{Q}}|=0$, $(\ee^{-tH+\theta(-1)^F\overline{Q}})^\sim
=-\ee^{-tH+\theta(-1)^F\overline{Q}}$ and that $\ee^{-tH+\theta(-1)^F\overline{Q}}$
is invertible as required. 
The effective BV bracket and Laplacian $(-,-)^0{}_{t\theta}$, $\varDelta^0{}_{t\theta}$ are then given by 
the accompanying deformations of $(-,-)_E$, $\varDelta_E$, viz 
\begin{align}
&(u,v)^0{}_{t\theta}=(u,v)_{\ee^{-tH+\theta(-1)^F\overline{Q}}}
=(u,v)_{\ee^{-tH}}+\theta(-1)^{|u|}(u,v)_{\overline{Q}\ee^{-tH}},
\vphantom{\Big]}
\label{freemod20/1}
\\
&\varDelta^0{}_{t\theta}=\varDelta_{\ee^{-tH+\theta(-1)^F\overline{Q}}}
=\varDelta_{\ee^{-tH}}+\theta\varDelta_{\overline{Q}\ee^{-tH}}  
\vphantom{\Big]}
\label{freemod15}
\end{align}
defined complying with \ceqref{nsumbvrg3}, \ceqref{nsumbvrg4}, respectively.

The RG flow $\varphi^0{}_{t\theta,s\zeta}$ of the free BV EFT is generated infinitesimally by the degree $0$, $-1$ 
vector fields $\langle x,H\ad_E x\rangle_E$, $\langle x,\overline{Q}\ad_E x\rangle_E$, 
\begin{align}
&\varphi^0{}_{t\theta,s\zeta}=
\exp\bigg(-\zeta\frac{1}{2}\langle x,\overline{Q}\ad_E x\rangle_E\bigg)\circ
\vphantom{\Big]}
\label{freemod16}
\\
&\hspace{3cm}\exp\bigg(\frac{t-s}{2}\langle x,H\ad_E x\rangle_E\bigg)
\circ\exp\bigg(\theta\frac{1}{2}\langle x,\overline{Q}\ad_E x\rangle_E\bigg).
\vphantom{\Big]}
\nonumber
\end{align}
Its logarithmic Jacobian again vanishes,
\begin{equation}
r_{\varphi^0 t\theta,s\zeta}=0. 
\label{freemod17}
\end{equation}
The pull--back action $\varphi^0{}_{t\theta,s\zeta}{}^*$ of $\varphi^0{}_{t\theta,s\zeta}$ on the function algebra 
$\Fun(E_0)$ is given by the right hand side of \ceqref{freemod17} with the three exponential factors
in reversed order upon regarding $\langle x,H\ad_E x\rangle_E$, $\langle x,\overline{Q}\ad_E x\rangle_E$
as derivations of 
$\Fun(E_0)$.  

The proof that $\varphi^0{}_{t\theta,s\zeta}$ is a BV RG flow follows the same lines as that of the 
corresponding property in the basic case. Relations \ceqref{bvfix11}--\ceqref{bvfix13} again hold trivially. 
The proof that the maps $\varphi^0{}_{t\theta,s\zeta}$ are canonical, that is that   
they intertwine the BV brackets $(-,-)^0{}_{t\theta}$ as in \ceqref{bvint9}, follows from the relation
\begin{equation}
(u,v)^0{}_{t\theta}=\varphi^0{}_{t\theta,00}{}^*(\varphi^0{}_{t\theta,00}{}^{-1*}u, \varphi^0{}_{t\theta,00}{}^{-1*}v)_E,
\label{nfreemod7}
\end{equation}
analogous to \ceqref{nfreemod1}, which can be shown again using \ceqref{sumbvrg15/1}. 
The vanishing of $r_{\varphi^0 t\theta,s\zeta}$ can be inferred by comparing the relation \pagebreak 
\begin{equation}
\varDelta^0{}_{t\theta}=\varphi^0{}_{t\theta,00}{}^*\varDelta_E\varphi^0{}_{t\theta,00}{}^{-1*},
\label{nfreemod8}
\end{equation}
analogous to \ceqref{nfreemod2} and proven again using \ceqref{sumbvrg15}, and \ceqref{bvint13}. 

In the extended theory, the appropriate enhancement of the basic theory BV EA 
is the two--parameter family of free actions $S^0{}_{t\theta}\in\Fun(E_0)$ given by 
\begin{equation}
S^0{}_{t\theta}=-\frac{1}{2}\langle x,Q \ee^{tH}x\rangle_E
-\theta\frac{1}{4}\langle x,\{\overline{Q},Q\} \ee^{tH}x\rangle_E,
\label{freemod21}
\end{equation}
where $\{\overline{Q},Q\}=\overline{Q}Q-Q\overline{Q}$. 
We are now going to show that $S^0{}_{t\theta}$ has the required properties of a BV RG EA. 

$S^0{}_{t\theta}$ obeys the extended BV quantum ME \ceqref{bveffact26}. This follows again 
from the identities \ceqref{sumbvrg25}, \ceqref{sumbvrg24} and from \ceqref{n2struct4} and \ceqref{n2struct7}
and the graded cyclic in\-variance of graded trace $\grtr$, 
\begin{align}
&\varDelta^0{}_{t\theta}S_{t\theta}+\frac{1}{2}(S_{t\theta},S_{t\theta})^0{}_{t\theta}
=-\frac{1}{2}\grtr(Q)+\frac{1}{2}\langle x,Q^2\ee^{tH}x\rangle_E
\vphantom{\Big]}
\label{nfreemod9}
\\
&+\theta\bigg[-\frac{1}{2}\grtr(\overline{Q}Q)+\frac{1}{4}\grtr(\{\overline{Q},Q\})
\vphantom{\Big]}
\nonumber
\\
&\hspace{2.5cm}+\frac{1}{2}\langle x, Q\overline{Q}Q\ee^{tH}x \rangle_E
-\frac{1}{2}\langle x, Q\{\overline{Q},Q\} \ee^{tH}x \rangle_E\bigg]=0.
\vphantom{\Big]}
\nonumber
\end{align}

Next, $S^0{}_t$ satisfies the RG flow relation \ceqref{nnbveffact7/2}.
We first note that 
\begin{equation}
S^0{}_{t\theta}=\varphi^0{}_{0\theta,00}{}^*S^0{}_{t0}, 
\label{nfreemod10}
\end{equation}
as can be proven easily using that 
$\varphi^0{}_{0\theta,00}=\exp\big(\theta\frac{1}{2}\langle x,\overline{Q}\ad_E x\rangle_E\big)$
by \ceqref{freemod16} and employing \ceqref{sumbvrg26}. Next, we note that \hphantom{xxxxxxxxxx}
\begin{equation}
S^0{}_{t0}=\varphi^0{}_{t0,s0}{}^*S^0{}_{s0}
\label{nfreemod11}
\end{equation}
by the same calculation which proves the analogous relation of the basic case detailed above, since
$S^0{}_{t0}=S^0{}_t$ and $\varphi^0{}_{t0,s0}=\varphi^0{}_{t,s}$. It follows by \ceqref{freemod16} that 
\begin{equation}
\varphi^0{}_{t\theta,s\zeta}{}^*S^0{}_{s\zeta}
=\varphi^0{}_{0\theta,00}{}^*\varphi^0{}_{t0,s0}{}^*\varphi^0{}_{0\zeta,00}{}^{-1*}\varphi^0{}_{0\zeta,00}{}^*S^0{}_{s0}=S^0{}_{t\theta}
\label{nfreemod12}
\end{equation}
as required. 

In the extended set-up, the RGE obeyed by $S^0{}_t=S^0{}_{t0}$ can be cast as  \pagebreak
\begin{equation}
\frac{d S^0{}_t}{dt}=-\varDelta_{\overline{Q}\ee^{-tH}}S^0{}_t-\frac{1}{2}(S^0{}_t,S^0{}_t)_{\overline{Q}\ee^{-tH}}
-\frac{1}{2}\grtr(\overline{Q}Q),
\label{freemod28}
\end{equation}
as is immediately verified using \ceqref{freemod21} together with the identities 
\ceqref{sumbvrg25}, \ceqref{sumbvrg24}. 

By the results of subsect. \cref{subsec:rgsusy}, we expect that 
the RGE \ceqref{freemod14} can be cast in the form \ceqref{bveffact27}.
Eq. \ceqref{freemod28} apparently deviates from \ceqref{bveffact27} by 
the sign of the first two terms. This mismatch is however only apparent.
As we shall argue in subsect. \cref{subsec:permod} below, 
the reduced infinitesimal generator and Jacobian of the BV flow,
$\bar\varphi^{0\bcdot}{}_t$ and $\bar r^\bcdot{}_{\varphi^0 t}$,  contain those very same terms with
coefficients such to produce the result shown.


\subsection{\textcolor{blue}{\sffamily Perturbative BV RG}}\label{subsec:permod}  

In this final subsection, we analyze the implications of the results found  in subsect. \cref{subsec:freemod}  
for the perturbative BV RG. 

In perturbation theory, one promotes the function algebra $\Fun(E_0)$ to the formal power series algebra
$\Fun(E_0)[[\hbar]]$ of the parameter $\hbar$ over $\Fun(E_0)$. Further, for fixed $t$, 
the full action $S_t\in\Fun(E_0)[[\hbar]]$ 
is rescaled by $\hbar^{-1}$. It is further assumed that $S_t$ satisfies 
the same BV quantum ME as the free action $S^0{}_t$, 
\begin{equation}
\hbar\varDelta_{\ee^{-tH}}S_t+\frac{1}{2}(S_t,S_t)_{\ee^{-tH}}=0.
\label{permod1}
\end{equation}
Aiming to a perturbative analysis of the BV RG, one splits the action $S_t$ as 
\begin{equation}
S_t=S^0{}_t+I_t,
\label{permod2}
\end{equation}
where $S^0{}_t$ 
is the free action \ceqref{freemod8} and 
$I_t\in\Fun(E_0)[[\hbar]]$ is an interaction term at least cubic in $x$ mod $\hbar$. 

A simple calculation shows that $I_t$ obeys the BV quantum ME
\begin{equation}
\hbar\varDelta_{\ee^{-tH}}I_t-\mathcal{Q}I_t+\frac{1}{2}(I_t,I_t)_{\ee^{-tH}}=0,
\label{permod3}
\end{equation}
where $\mathcal{Q}$ is the degree $1$ first order differential operator defined by 
\begin{equation}
\mathcal{Q}u=\langle x,Q(x,u)_E\rangle_E \pagebreak 
\label{permod4}
\end{equation}
acting on $\Fun(E_0)[[\hbar]]$. 
Eq. \ceqref{permod3} is precisely of the same form as the ME for $I_t$ obtained by Costello in 
ref. \ccite{Costello:2007ei}. 

The next problem we have to address is determining the BV RG flow 
of the full action $S_t$. We assume as a working hypothesis that 
this is governed by the Polchinski's RGE with seed
action $S^0{}_t$ reducing to the free RGE \ceqref{freemod28} in the limit of vanishing $I_t$ is 
\begin{equation}
\frac{dS_t}{dt}=\hbar\varDelta_{\overline{Q}\ee^{-tH}}(S_t-2S^0{}_t)
+\frac{1}{2}(S_t,S_t-2S^0{}_t)_{\overline{Q}\ee^{-tH}}-\frac{\hbar}{2}\grtr(\overline{Q}Q).
\label{permod7}
\end{equation}
The last term may be absorbed by adding a constant term $\frac{t}{2}\grtr(\overline{Q}Q)$ to $S_t$
and for this reason is usually neglected, but we shall keep it for the time being. 
Next, we are going to argue that \ceqref{permod7} is the correct full RGE. 

By the perturbative decomposition \ceqref{permod2}, the RGE \ceqref{permod7} can be cast as 
\begin{equation}
\frac{dI_t}{dt}=\hbar\varDelta_{\overline{Q}\ee^{-tH}}I_t+\frac{1}{2}(I_t,I_t)_{\overline{Q}\ee^{-tH}}.
\label{permod8}
\end{equation}
By \ceqref{freemod20/1}, \ceqref{freemod15}, and the standard identity 
$\ee^{-u}\varDelta\ee^u=\varDelta u+\frac{1}{2}(u,u)$ 
of BV theory, \ceqref{permod8} can be written more compactly as 
\begin{equation}
\frac{dI_t}{dt}=\hbar\varDelta_{\overline{Q}\ee^{-tH}}e^{I_t/\hbar}. 
\label{permod10}
\end{equation}
The formal solution of \ceqref{permod10} is 
\begin{equation}
\ee^{I_t/\hbar}=\exp\bigg(\hbar\int_s^t d\tau \varDelta_{\overline{Q}\ee^{-\tau H}}\bigg)\ee^{I_s/\hbar}.
\label{permod11}
\end{equation}
This relation constitutes one of the characterizing property of $I_t$ in the analysis of ref. \ccite{Costello:2007ei}. 

The RGE \ceqref{permod7} can be reshaped in the form
\begin{equation}
\frac{dS_t}{dt}=\hbar\varDelta_{\overline{Q}\ee^{-tH}}S_t+\frac{1}{2}(S_t,S_t)_{\overline{Q}\ee^{-tH}}
+\bar\varphi^\bcdot{}_tS_t+\hbar \bar r^\bcdot{}_{\varphi t},
\label{permod12}
\end{equation}
where $\bar\varphi^\bcdot{}_t$ and $\bar r^\bcdot{}_t,$ are the degree $0$ derivation and element of $\Fun(E_0)$
\begin{align}
&\bar\varphi^\bcdot{}_t=-\ad_{\overline{Q}\ee^{-tH}}S^0{}_t,
\vphantom{\Big]}
\label{permod13}
\\
&\bar r^\bcdot{}_{\varphi t}=-2\varDelta_{\overline{Q}\ee^{-tH}}S^0{}_t-\frac{1}{2}\grtr(\overline{Q}Q).
\vphantom{\Big]}
\label{permod14}
\end{align}
The reason for this rewriting of \ceqref{permod7} is that 
\ceqref{permod12} has the same form as the RGE \ceqref{bveffact27} of the extended RG set--up 
of subsect. \cref{subsec:rgsusy}. \ceqref{permod12} may provide in this way useful indications about 
the underlying BV RG flow.

In the extended RG set--up, the full BV RG EA $S_t$ has degree $-1$ partner
$S^\star {}_t$ and $S_t$ and $S^\star {}_t$ obey eq. \ceqref{bveffact26}. Though it is not obvious 
that such partner exists, let us assume that it does. 
Eq. \ceqref{bveffact26} would then read 
\begin{equation}
\hbar \varDelta_{\ee^{-tH}}S^\star {}_t+(S_t,S^\star {}_t)_{\ee^{-tH}}
=\hbar \varDelta_{\overline{Q}\ee^{-tH}}S_t+\frac{1}{2}(S_t,S_t)_{\overline{Q}\ee^{-tH}}.
\label{permod14/1} 
\end{equation}   
Correspondingly,  the RGE \ceqref{permod12} would take the form 
\begin{equation}
\frac{dS_t}{dt}=\varphi^\bcdot{}_tS_t+\hbar r^\bcdot{}_{\varphi t},
\label{permod14/2}
\end{equation}
where $\varphi^\bcdot{}_t$ and $r^\bcdot{}_t$ are the degree $0$ derivation and element of $\Fun(E_0)$ 
\begin{align}
&\varphi^\bcdot{}_t=-\ad_{\ee^{-tH}}S^\star {}_t+\bar\varphi^\bcdot{}_t,
\vphantom{\Big]}
\label{permod16}
\\
&r^\bcdot{}_{\varphi t}=\varDelta_{\ee^{-tH}}S^\star {}_t+\bar r^\bcdot{}_{\varphi t}.
\vphantom{\Big]}
\label{permod17}
\end{align}
It is reasonable to hypothesize that, in analogy to $S_t$, $S^\star {}_t$ splits as 
\begin{equation}
S^\star {}_t=S^{0\star}{}_t+I^\star {}_t, 
\label{permod15}
\end{equation}
where $S^{0\star}{}_t$ 
is the degree $-1$ partner of the free action $S^0{}_t$ given by the second term in the right hand side of
\ceqref{freemod21} and $I^\star {}_t\in\Fun(E_0)[[\hbar]]$ is similarly a partner 
of the interaction term $I_t$ likewise cubic in $x$ mod $\hbar$. 

Under the above assumptions, $\varphi^\bcdot{}_t$ and $r^\bcdot{}_{\varphi t}$ are given by  
\begin{align}
&\varphi^\bcdot{}_t=\frac{1}{2}\langle x, H\ad_E x\rangle_E -\ad_{\ee^{-tH}}I^\star {}_t,
\vphantom{\Big]}
\label{permod18}
\\
&r^\bcdot{}_{\varphi t}=\varDelta_{\ee^{-tH}}I^*{}_t.
\vphantom{\Big]}
\label{permod19}
\end{align}
We obtain these identities by substituting relations \ceqref{permod13}, \ceqref{permod14}
and \ceqref{permod15} into \ceqref{permod16}, \ceqref{permod17}, inserting the explicit expressions
of $S^0{}_t$ and $S^{0\star }{}_t$ appearing in \ceqref{freemod21} into the resulting expressions
and simplifying. It turns out that, under the assumptions made, 
$\varphi^\bcdot{}_t$ and $r^\bcdot{}_t$ are the infinitesimal generator and logarithmic Jacobian 
of a BV RG flow $\varphi_{t,s}$. $\varphi_{t,s}$ is the flow generated infinitesimally by the vector field
$\varphi^\bcdot{}_t$. The conformality of $\varphi_{t,s}$ is a consequence of the equation
\begin{equation}
\varphi^\bcdot{}_t(u,v)_{\ee^{-tH}}=(\varphi^\bcdot{}_tu,v)_{\ee^{-tH}}
+(u,\varphi^\bcdot{}_tv)_{\ee^{-tH}}+\frac{d}{dt}(u,v)_{\ee^{-tH}},
\label{npermod1}
\end{equation}
which can be verified using \ceqref{sumbvrg15/1}. The expression \ceqref{permod19} of the logarithmic Jacobian 
of $\varphi_{t,s}$ can be inferred by its satisfying  the equation 
\begin{equation}
\frac{d}{dt}\varDelta_{\ee^{-tH}}-[\varphi^\bcdot{}_t,\varDelta_{\ee^{-tH}}]+\ad_{\ee^{-tH}}r^\bcdot{}_{\varphi t}=0
\label{npermod2}
\end{equation}
following from \ceqref{sumbvrg15}. By \ceqref{permod14/2}, then, 
$S_t$ is a BV RG EA flowing according to  $\varphi_{t,s}$. 

The infinitesimal generator and Jacobian $\varphi^{0\bcdot}{}_t$ and $r^\bcdot{}_{\varphi^0 t}$ 
of the special free BV RG flow $\varphi^0{}_{t,s}$ of eq. \ceqref{freemod2} are \hphantom{xxxxxxxx} 
\begin{align}
&\varphi^{0\bcdot}{}_t=\frac{1}{2}\langle x, H\ad_E x\rangle_E, 
\vphantom{\Big]}
\label{permod24}
\\
&r^\bcdot{}_{\varphi^0 t}=0. 
\vphantom{\Big]}
\label{permod25}
\end{align}
Comparing \ceqref{permod18}, \ceqref{permod19} with \ceqref{permod24}, \ceqref{permod25}, it appears that 
the full BV RG flow $\varphi_{t,s}$ differs form its free counterpart $\varphi^0{}_{t,s}$ 
by an amount determined by 
$I^*{}_t$. As this latter, so, $\varphi_{t,s}$ has a formal perturbative expansion. 

The importance of the BV RG EA partner's interaction term $I^*{}_t$ should by now be clear, 
although presently we cannot prove its existence in general. On account of 
\ceqref{permod2}, \ceqref{permod15}, $I^*{}_t$ is related 
to BV RG EA interaction term $I_t$ through eq. \ceqref{permod14/1}. An elementary calculations 
shows that 
\begin{equation}
\hbar \varDelta_{\ee^{-tH}}I^*{}_t-\mathcal{Q}I^*{}_t+(I_t,I^*{}_t)_{\ee^{-tH}}
=\hbar \varDelta_{\overline{Q}\ee^{-tH}}I_t-\frac{1}{2}\mathcal{H}I_t+\frac{1}{2}(I_t,I_t)_{\overline{Q}\ee^{-tH}},
\label{permod3/1}
\end{equation} 
where $\mathcal{Q}$ is defined in \ceqref{permod4} and $\mathcal{H}$ is the degree $0$ first order differential operator
\begin{equation}
\mathcal{H}u=\langle x,H(x,u)_E\rangle_E.
\label{permod4/1}
\end{equation}
Eq. \ceqref{permod3/1} is a further quantum ME involving simultaneously $I_t$, $I^\star{}_t$ 
and complementing the ME \ceqref{permod3}. It can be recast in an alternative form using the RGE 
\ceqref{permod8}.

\vfil\eject


\section{\textcolor{blue}{\sffamily Outlook, beyond renormalization group}}\label{sec:beyond}

In this final section, we speculate about possible applications of the BV RG framework
we have described in depth in the main body of the paper. 

In physical applications of RG theory, much effort is devoted to the derivation of
the RGEs of the couplings of the basic fields in the EA and the computation of the appended
beta functions. These RGEs are implicit in the RGE of the EA. It would be interesting to devise 
systematic methods to obtain them within the BV RG framework of this paper. See ref.
\ccite{Elliott:2017ahb} for an alternative approach to this problem.

The connection between the weak coupling limit of the RG flow of 
two--dimensional non linear sigma models, originally studied by Friedan in 
\ccite{Friedan:1980th,Friedan:1980jf,Friedan:1980jm},
and the Ricci flow, introduced by Hamilton in  \ccite{Hamilton:1982rc}, 
is by now established \ccite{Carfora:2010iz,Grady:2017ns}. 
Ricci flow has has played a pivotal role in important developments of geometric analysis, 
such as  Perelman's proof of Thurston's geometrization program 
\ccite{Thurston:1982zz,Thurston:1997bk} for three--manifolds and the related Poincar\'e conjecture 
\ccite{Perelman:2006un,Perelman:2006up,Perelman:2003uq}. 

The Alexandrov--Kontsevich--Schwartz--Zaboronsky (AKSZ) formulation of the BV quantization 
scheme \ccite{Alexandrov:1995kv} is a general framework for the construction of the BV MA
of a broad class of sigma models, at least in the semiclassical limit. It has had a wide range of 
field theoretic applications with remarkable mathematical ramifications. 
See ref. \ccite{Ikeda:2012pv} for a recent review of the AKSZ approach. 

In the AKSZ approach, the fulfilment of the BV ME by the MA of a sigma model, and so its ultimate 
quantum consistency, rests on relevant features of the model's target space geometry. The AKSZ approach 
can therefore be employed to construct for any given target geometry a canonical sigma model capable 
of probing it. 
It is conceivable that a carefully designed BV RGE for the model may yield a 
flow equation analogous to Ricci flow potentially useful for the study of important geometrical and topological issues of 
the target manifold. This remains at the moment an unexplored possibility.

\vfil\eject

\noindent
\textcolor{blue}{\sf Acknowledgements}. 
The author thanks G. Velo for useful discussions.
The author acknowledges financial support from INFN Research Agency
under the provisions of the agreement between Bologna University and INFN. 
The author also thanks for hospitality 
the Instituto Superior T\'ecnico of Lisbon, where a part of this work was done
in occasion of the conference ``Higher Structures 2017''.

\vfil\eject

\end{document}